\documentclass[prd,aps,twocolumn,a4paper,showkeys,nofootinbib]{revtex4-1}

\usepackage[colorlinks=true,backref=true,linkcolor=blue,linktocpage=true,anchorcolor=black,citecolor=blue,filecolor=black,menucolor=black,pagecolor=black,urlcolor=blue]{hyperref}
\usepackage{graphicx,psfrag}
\usepackage{mathrsfs}
\usepackage{amsmath,amsfonts,amssymb}
\usepackage{multirow}
\usepackage{comment}
\usepackage{bm}
\usepackage{paralist}
\usepackage{ulem}
\usepackage{xcolor}
\usepackage{enumitem}
\usepackage{bm}		
\usepackage[T1]{fontenc}

\usepackage{pifont} 
\newcommand{\be}{\begin{equation}}
\newcommand{\ee}{\end{equation}}
\newcommand{\bea}{\begin{eqnarray}}
\newcommand{\eea}{\end{eqnarray}}
\newcommand{\bel}{\begin{align}}
\newcommand{\eel}{\end{align}}

\newcommand{\ord}{\mathcal{O}}
\newcommand{\f}{\frac}

\newcommand{\el}{\ell}
\newcommand{\nn}{\nonumber}
\newcommand{\mbf}[1]{\mathbf{#1}}
\newcommand{\LOhat}{\!\hat{\,\mathbf{L}}_0}
\newcommand{\Lhat}{\!\hat{\,\mathbf{L}}_\text{N}}
\newcommand{\Jhat}{\hat{\mathbf{J}}_\text{N}}
\newcommand{\Sa}{\mbf{S}_1}
\newcommand{\Sb}{\mbf{S}_2}
\newcommand{\Lhatdot}{\dot{\hat{\,\mathbf{L}}}_\text{N}}
\newcommand{\LhatdotNew}{\dot{\hat{\,\mathbf{L}}}_{\text{N},\perp}}
\newcommand{\Sadot}{\dot{\mbf{S}}_1}
\newcommand{\Sbdot}{\dot{\mbf{S}}_2}
\newcommand{\JN}{\mbf{J}_\text{N}}
\newcommand{\LN}{\mbf{L}_\text{N}}
\newcommand{\Jivec}{\mbf{J}_0}
\newcommand{\Livec}{\mbf{L}_0}

\newcommand{\fgw}{f}
\newcommand{\fgwi}{f_0}
\newcommand{\J}{\mbf{J}}
\renewcommand{\S}{\mbf{S}}
\renewcommand{\L}{\mbf{L}}

\newcommand{\la}{\lambda}

\newcommand{\lhatp}{{\bm{\lambda}}}
\newcommand{\elhatp}{{\bm{\ell}}}
\newcommand{\nhatp}{{\bm{n}}}
\newcommand{\etal}{{\it et al.}\,}

\newcommand{\Sperp}{\chi_{\perp,\text{max}}}
\newcommand{\nNR}{200}
\newcommand{\nST}{50}
\newcommand{\nTot}{1230}

\def\i{{\rm i}}

\def\Msun{M_{\odot}}
\def\Mtot{M_{\text{tot}}}
\def\GMc2{G M_{\odot} c^{-2}}

\def\M{\mathcal{M}}
\def\Mbar{\bar{\M}}

\def\M{{\cal M}}

\def\Msun{M_\odot}

\def\TEOBResumS{\texttt{TEOBResumS}}
\def\TEOBResumP{\texttt{TEOBResumSP}}
\def\TEO{\texttt{TEOB}}
\def\TEOB{\TEOBResumP}
\def\TEOBPv0{\texttt{TEOBResumPv0}}

\def\PhenomPv3HM{\texttt{PhenomPv3HM}}
\def\PHM{\texttt{IMRPhenomPv3HM}}

\def\STT{{\texttt{SpinTaylorT4}}}

\def\SEOBNR{{\texttt{SEOBNRv4PHM}}}
\def\SEOB{{\texttt{SEOB}}}
\def\NRsurP{{\texttt{NRSur7dq4}}}
\def\NRsur{{\texttt{NRSurHyb3dq8}}}
\def\NR{{\texttt{NRSur}}}
\def\Phn{{\texttt{Pv3HM}}}

\DeclareSymbolFontAlphabet{\mathrsfs}{rsfs}
\DeclareMathAlphabet{\mathcal}{OMS}{cmsy}{m}{n}
\DeclareSymbolFontAlphabet{\mathrsfs}{rsfs}
\DeclareMathAlphabet\mathbfcal{OMS}{cmsy}{b}{n}

\usepackage{color}
\definecolor{cyan}{rgb}{0,0.9,0.9}
\definecolor{orange}{rgb}{0.9,0.5,0}
\definecolor{magenta}{rgb}{1,0,1}
\definecolor{purple}{rgb}{0.8,0.4,0.8}
\definecolor{gray}{rgb}{0.8242,0.8242,0.8242}
\definecolor{dodgerblue}{rgb}{0.12, 0.56, 1.0}

\newcommand{\todo}[1]{\textcolor{orange}{\texttt{TODO: #1}}}

\begin{document}

\title{
 A hybrid post-Newtonian -- effective-one-body scheme for
  spin-precessing compact-binary waveforms
  up to merger
  }

\author{Sarp \surname{Akcay}$^{1,2}$}
\author{Rossella \surname{Gamba}$^{2}$}
\author{Sebastiano \surname{Bernuzzi}${}^{2}$}

\affiliation{${}^1$University College Dublin, D14, Dublin, Ireland}
\affiliation{${}^2$Theoretisch-Physikalisches Institut, Friedrich-Schiller-Universit{\"a}t Jena, 07743, Jena, Germany}

\normalem 

\begin{abstract}
 We introduce \texttt{TEOBResumSP}: an efficient yet accurate hybrid scheme for 
 generating gravitational waveforms from 
 spin-precessing compact binaries. 
 The precessing waveforms are generated
 via the established technique of Euler rotating 
 aligned-spin (non-precessing) waveforms 
 from a precessing frame to an inertial frame.
 We employ the effective-one-body approximant \texttt{TEOBResumS} to generate the aligned-spin waveforms. 
 We obtain the Euler angles by
 solving the post-Newtonian precession equations expanded to
 (next-to)$^4$ leading (second post-Newtonian) order.
 Current version of \texttt{TEOBResumSP} produces precessing waveforms
 through the inspiral phase up to the onset of the merger.
 We compare \texttt{TEOBResumSP} to current state-of-the-art precessing approximants 
 \texttt{NRSur7dq4}, \texttt{SEOBNRv4PHM}, and \texttt{IMRPhenomPv3HM} in terms of frequency-domain matches of the
 $\el=2$ gravitational-wave strain for \nNR{} cases of precessing compact binary inspirals
 with orbital inclinations up to 90 degrees, mass ratios up to four, and
 the effective precession parameter $\chi_p$ up to 0.75.
 We further provide an extended comparison with
 \texttt{SEOBNRv4PHM} involving 1030 more inspirals
 with $\chi_p$ ranging up to one and mass ratios up to 10. 
 We find that 91\% of the \texttt{TEOBResumSP}-\texttt{NRSur7dq4} matches,
 85\% of the \texttt{TEOBResumSP}-\texttt{SEOBNRv4PHM} matches, 
 and 77\% of the \texttt{TEOBResumSP}-\texttt{IMRPhenomPv3HM} matches 
 are greater than $0.965$.
 Most of the significant disagreements occur for large mass ratios and $\chi_p \gtrsim 0.6$.
 We identify the mismatch of the \emph{non}-precessing $(2,1)$ mode as one of the leading causes of disagreements.
 We also introduce a new parameter, $\Sperp$, 
 to measure the strength of precession
 and hint that the strain mismatch between the above waveform approximants
 shows an exponential dependence on $\Sperp$ though this requires further study.
 Our results indicate that \texttt{TEOBResumSP} is on its way to becoming a robust precessing approximant
 to be employed in the parameter estimation of generic-spin
 compact binaries.
 \end{abstract}


\maketitle


\section{Introduction}\label{Sec:introduction}
Gravitational-wave events have become routine in observational astronomy:
the Advanced LIGO \cite{TheLIGOScientific:2014jea}-Virgo \cite{TheVirgo:2014hva} interferometers 
detected at least ten binary black hole coalescences 
and one binary neutron star merger during the first and second observing runs
\cite{Abbott:2016blz, TheLIGOScientific:2017qsa, Abbott:2016nmj, Abbott:2017gyy, Abbott:2017oio, Abbott:2017vtc, LIGOScientific:2018mvr, 
Venumadhav:2019tad,Venumadhav:2019lyq, Nitz:2018imz, Nitz:2019hdf}.
The third observing run began on 1 April 2019 and delivered by its 
[premature] end a year later the second binary neutron star merger \cite{Abbott:2020uma}, 
two binary black hole mergers with significant mass asymmetry \cite{LIGOScientific:2020stg,Abbott:2020khf},
the second possibly involving the most massive neutron star discovered yet,
another black hole merger leading to the formation of an intermediate mass 
black hole \cite{Abbott:2020tfl},
and additionally more than four dozen triggers with false alarm rates of less than one per year \cite{gracedb}.
A significant fraction of these triggers turned out to be genuine gravitational-wave events
caused by the inspiral and merger of stellar mass compact objects \cite{Abbott:2020niy}.

The properties of the compact objects such as masses
and spins can be obtained via parameter estimation studies that are conducted on a sufficiently ``cleaned'' version of the relevant
segment of the detector data.
This requires a large set of ``realistic'' theoretical gravitational waveform templates which can be cross-correlated with the data.
For stellar-mass compact binary systems, 
there are four main approaches to generating the theoretical
gravitational waves (GWs) resulting from compact binary inspirals: 
post-Newtonian theory \cite{Blanchet:2013haa}, numerical relativity \cite{Gourgoulhon:2007ue, Centrella:2010mx},
effective-one-body theory \cite{Buonanno:1998gg, Buonanno:2000ef}, and
phenomenological template construction \cite{Ajith:2007qp, Ajith:2007kx}.
More recently, there has also been an emergence of surrogate methods
which we discuss below.

Post-Newtonian (PN) theory employs a large-separation (weak-field) expansion to the Einstein field equations.
Current PN technology for the evolution of quasi-circular inspirals is at the 3.5PN level with partial higher-order
PN information available \cite{Bernard:2017ktp, Messina:2019uby}. 
As PN information is fully analytical, 
the resulting waveforms can be evaluated very quickly. 
Consequently, the LIGO-Virgo Collaboration (LVC)
has at its disposal a plethora of PN-based Taylor waveform approximants 
summarized in Ref.~\cite{Buonanno:2009zt}.
As PN theory is valid in the weak-field, adiabatic regime,
these approximants are appropriate for modelling only the inspiral phase and extracting the chirp mass \cite{Damour:2012yf,Abbott:2020uma},
with possibly \texttt{PhenSpinTaylorRD} \cite{Sturani:2010yv}
as an exception, which is a hybrid model that matches spinning PN inspiral waveforms
and fits to NR ringdown waveforms.

Binary black hole (BBH) systems are more massive, thus transit through the LIGO-Virgo detection bandwidth 
much more quickly than binary neutron stars (BNSs), e.g., GW150914 lasted less than 20 milliseconds \cite{Abbott:2016blz} whereas GW170817 lasted nearly a minute \cite{TheLIGOScientific:2017qsa}.
As such, we can detect only the last few dozen cycles of their GWs before merger. Such GWs are generated 
in the very-strong-gravity regime where PN approximation is not reliable.
This is the domain of numerical relativity (NR). Since the breakthroughs of 2005 \cite{Pretorius:2005gq, Campanelli:2005dd, Baker:2005vv},
it has become routine to evolve strongly gravitating spacetimes 
of compact binary mergers on large computing clusters.
There are now several NR catalogs containing thousands of simulations of compact binary inspirals
\cite{Mroue:2013xna, Boyle:2019kee, SXS:catalog, Jani:2016wkt, GA_Tech:catalog, RIT:catalog, Healy:2017psd, Dietrich:2018phi, CoRe:catalog}.
Of these, the most comprehensive is the 2019 SXS catalog which contains 2018 simulations of
precessing systems with the dimensionless Kerr spin parameter up to 0.998 \cite{Boyle:2019kee}.

As the number of NR simulations increased, it became possible to build
hybrid (phenomenological) waveform models
by matching PN inspiral waveforms and fits to NR waveforms.
The initial model, \texttt{PhenomA} \cite{Ajith:2007qp, Ajith:2007kx}, 
combined the \texttt{TaylorT1} PN waveform model
with a two-dimensional fit to a set of non-precessing NR simulations.
The model was steadily improved through versions \texttt{B} \cite{Ajith:2009bn},
\texttt{C} \cite{Santamaria:2010yb}, and \texttt{D} \cite{Husa:2015iqa, Khan:2015jqa}.
Specific models were then developed for binary neutron stars
(\texttt{PhenomD\_NRTidal} \cite{Dietrich:2017aum, Dietrich:2019kaq}),
higher modes (\texttt{PhenomHM} \cite{London:2017bcn}),
and spin precession (\texttt{PhenomP} \cite{Hannam:2013oca, Schmidt:2014iyl}).
Subsequently, the non-precessing models have gone through several upgrades
\cite{Pratten:2020fqn, Garcia-Quiros:2020qpx, Garcia-Quiros:2020qlt},
just as the precessing ones have \cite{Chatziioannou:2017tdw, Khan:2018fmp, Khan:2019kot, Pratten:2020ceb, Estelles:2020osj}
with \texttt{IMRPhenomPv3HM} \cite{Khan:2019kot} being
employed in the analysis of the most recent GW events.
Note that the recent \texttt{IMRPhenomX} family match a mix of EOB, PN
waveforms with NR \cite{Pratten:2020fqn, Garcia-Quiros:2020qpx, Pratten:2020ceb}. 
Phenom models generate frequency-domain waveforms with the
corresponding time-domain waveforms obtained by inverse fast Fourier transforms,
only exception being \texttt{IMRPhenomTP} \cite{Estelles:2020osj} which is 
a direct time-domain construction.
Since GW data analysis is performed in the frequency-domain
and as Phenom waveforms are fast to generate,
the Phenom family has become one of the most commonly used set of waveform approximants
in the parameter estimation of GW events as well as in other areas of
GW science where fast, reliable waveforms are required. 

The effective-one-body (EOB) approach bridges PN theory and NR.
It maps the two-body PN motion to a geodesic motion in an effective spacetime
via a deformation performed in terms of the symmetric mass ratio \cite{Buonanno:1998gg, Buonanno:2000ef}.
In its core, EOB contains an effective Hamiltonian 
for aligned-spin systems,
which resums the PN series 
in a suitable way to better capture the effects of the strong-field regime \cite{Damour:2009kr}. 
The inspiral is driven by a specially factorized/resummed radiation-reaction force \cite{Damour:2014sva}.
The resulting multipolar gravitational waveforms are also written in a factorized form
\cite{Damour:2008gu, Pan:2010hz}. 
The analytical EOB model is further supplemented with input from non-precessing NR simulations, 
thus extending the EOB evolution through the merger and, if it exists, ringdown stages.
These so-called EOBNR models \cite{Damour:2014yha, Bohe:2016gbl, Nagar:2017jdw}
have been incorporated into several waveform approximants
\cite{Purrer:2014fza, Babak:2016tgq, Cotesta:2018fcv, Dietrich:2019kaq, Nagar:2018zoe}
that are used for parameter estimation studies of LIGO-Virgo GW events.
The main advantages of employing EOB-based waveform approximants for parameter estimation are that they 
\begin{inparaenum}[(i)]
     \item push the validity of the model beyond the PN weak-field regime
       \item can be extended to the full parameter space, and 
     \item are much faster to evolve than NR simulations. 
\end{inparaenum}
EOB models can also accurately
model binary neutron star coalescences from low frequencies and up to
merger \cite{Damour:2009wj, Damour:2012yf, Bernuzzi:2012ci, 
Lackey:2013axa, Bernuzzi:2014owa, Hinderer:2016eia, Steinhoff:2016rfi,
Dietrich:2018uni, Akcay:2018yyh, Lackey:2018zvw, Matas:2020wab},
thus offering a viable alternative to 
PhenomTidal models
\cite{Husa:2015iqa, Khan:2015jqa, Dietrich:2019kaq, Thompson:2020nei},
or to PN-based tidal models
(e.g., \texttt{TaylorF2} with tides up to 7.5PN order 
\cite{Damour:2012yf, Vines:2011ud, Henry:2020ski})
In short, EOB can provide NR-PN-faithful waveforms for parameter estimation studies of both long and short 
inspiral-merger-ringdown signals, and for extracting information about tides. 

Although it has thus far been very difficult to distinguish the effects of precession on the
gravitational waves from the few dozen sources hitherto detected,
there are at least four GW events for which it has been inferred 
that the pre-merger binary components have nonzero spin.
These are  GW151226 \cite{Abbott:2016nmj}, where at least one black hole has dimensionless spin $> 0.28$ \cite{LIGOScientific:2019fpa},
GW170729 \cite{LIGOScientific:2018mvr, Chatziioannou:2019dsz}, 
where at least one black hole has dimensionless spin $> 0.27$ \cite{LIGOScientific:2019fpa},
GW190412 where either the primary \cite{LIGOScientific:2020stg} 
or the secondary \cite{Mandel:2020lhv} has positive dimensionless spin
depending on the priors used, and GW190521 with both black holes
having dimensionless spins $>0.5$ \cite{Abbott:2020tfl, Abbott:2020mjq}.
There are two additional events, GW170121 and GW170403, that seem to have at least
one \emph{anti-aligned} spinning component \cite{Venumadhav:2019lyq}\footnote{These events were discovered discovered by groups
outside of the LIGO-Virgo Collaboration who additionally 
reported several more GW events \cite{Venumadhav:2019lyq, Nitz:2018imz, Nitz:2019hdf}.}.

In binaries containing spinning black holes and/or millisecond pulsars, 
the spin-orbit and the spin-spin interactions contribute significantly to 
the phase and modulate distinguishably the amplitude of the emitted GWs. 
For example, there are more than 20 precession cycles contributing to the 
phasing of the GWs for a BNS with total mass of 3\,$M_\odot$ inspiralling 
from 30\,Hz \cite{Apostolatos:1994mx}.
Therefore, given that the required relative phase errors of the theoretical waveform 
templates must be $\lesssim 5\times 10^{-4}$ to avoid waveform systematics 
with Advanced LIGO-Virgo design sensitivity \cite{Purrer:2019jcp},
the templates must incorporate the effects of precession.
Neglecting precession for high-mass ratio binaries can cause event rate 
losses of $\sim 15\%$ and as high as 25\% - 60\% for the worst 
cases \cite{Harry:2016ijz,Bustillo:2016gid}.
For the third generation detectors such as the Einstein Telescope and
Cosmic Explorer, these  errors will need to be
$\lesssim 10^{-6}$ which will be a tremendous challenge as
the EOB, PN, and Phenom template families
will need to be improved by three orders of magnitude
while NR errors will need to be reduced by
at least an order of magnitude \cite{Purrer:2019jcp}.

There has been a dedicated and an ever-increasing effort to produce accurate gravitational waveforms from precessing compact binary systems. 
Initial developments were made in post-Newtonian theory \cite{Barker:1979, Thorne:1984mz} 
after the pioneering work of Mathisson, Papapetrou, and Dixon (MPD)
on the motion of spinning test particles in curved spacetimes \cite{Mathisson:1937zz,Papapetrou:1951pa,Dixon:1970zza}.
There are now several waveform approximants available for
  precessing spin analysis (and implemented in the 
LIGO Algorithm Library (\texttt{LAL}) \cite{lal_approximants}),
which are: 
\begin{inparaenum}[(i)]
  \item $\mathtt{SpinTaylorT}$ class of approximants 
which employ 1.5PN analytical expressions of Ref.~\cite{Arun:2008kb} for the waveform harmonic modes as functions of the spherical angles of the Newtonian orbital angular momentum vector.
  \item $\mathtt{IMRPhenomP}$ class of approximants 
  which transform non-precessing Phenom waveforms into precessing ones 
  using Euler rotations for which the angles are obtained from the
  PN spin precession equations 
  \cite{Hannam:2013oca, Schmidt:2014iyl, Khan:2018fmp, Khan:2019kot, Pratten:2020ceb, Estelles:2020osj}.
In particular, Ref.~\cite{Hannam:2013oca} showed that 
``the essential phenomenology of the seven-dimensional parameter space of
binary configurations'' can be modeled using just three parameters.
  \item $\mathtt{SEOBNR}$ class of approximants \cite{Pan:2013tva, Babak:2016tgq, Ossokine:2020kjp} which evolve the EOB dynamics and precession equations as a coupled system to determine the Euler angles for the
  rotation of the non-precessing waveform modes.
  \item \texttt{NRSur} class which are surrogate waveform models
  in which the surrogate is trained using large sets of precessing
  NR waveforms that are Euler-rotated to a certain non-inertial co-orbital frame.
\end{inparaenum}
With the exception of the \texttt{NRSur} family,
the above-listed approximants solve the same precession equations, albeit truncated at different PN orders or suitably incorporated into a particular EOB Hamiltonian. 
The solutions to the precession equations are then translated into the spherical angles of
the Newtonian orbital angular momentum. 
The precessing waveforms are constructed either via the analytical 
1.5\,PN expressions of Ref.~\cite{Arun:2008kb} (only for the \texttt{SpinTaylorT} family)
or by using the so-called {\it twist} method of Ref.~\cite{Schmidt:2010it} which
is what concerns us in this article so we provide some details next.

The seeds of the twist method were sown in Ref.~\cite{Apostolatos:1994mx}\footnote{Though, 
the frame rotation mentioned in App.~B of Ref.~\cite{Cutler:1994ys}
could possibly be taken as a hint of the twist method.},
where it was identified that the waveform phase can be decomposed into 
a non-modulating main carrier phase and a modulation term due to precession.
Ref.~\cite{Buonanno:2002fy}
used this decomposition to construct waveform templates
with the unmodulated carrier phase given by nonspinning frequency domain fits 
to the full 2\,PN phase.
It was later shown in Refs.~\cite{Schmidt:2010it, Schmidt:2012rh}
that the correct unmodulated carrier phase is given by the non-precessing,
but spinning phase.
Ref.~\cite{Buonanno:2002fy} also introduced a special non-inertial frame, 
called the precessing frame, in which 
the orbital phase agreed with the PN orbital phase of a non-spinning system. 
In other words, the modulations in the gravitational waveform phase due to precession
factored out. 
Subsequently, Ref.~\cite{Gualtieri:2008ux} obtained rigorous expressions for
the transformation of waveform multipoles under rotations, which were then employed by
Ref.~\cite{Campanelli:2008nk}  in order to
generate precessing post-Newtonian waveforms to compare with their numerical results.

A crucial step toward obtaining full (inspiral-merger-ringdown) precessing 
waveforms was taken by Ref.~\cite{Schmidt:2010it}
which employed a time-dependent frame rotation of the harmonic modes of the
Weyl scalar $\Psi_4$ into the ``quadrupole-aligned'' (QA)
frame defined by the direction toward which the amplitudes of the
$(2,\pm 2)$ modes are maximized, which turned out to coincide with 
the instantaneous direction of the total orbital angular momentum vector.
Ref.~\cite{Schmidt:2012rh} used this frame rotation on the $\el=2$
modes of the gravitational waveform and demonstrated that the model 
is better than 99\% accurate.
Ref.~\cite{OShaughnessy:2011pmr} introduced a frame similar to the QA frame
by  equating the radiation axis with the eigenvector of the rotation group generators which
had the largest absolute eigenvalue. 
Subsequently, Ref.~\cite{Boyle:2011gg} demonstrated that the special frames
of Refs.~\cite{Schmidt:2010it, Schmidt:2012rh} and Ref.~\cite{OShaughnessy:2011pmr} 
are the same if one includes only the $(2,\pm2)$ modes in the $m$-mode sum.
Additionally, Ref.~\cite{Boyle:2011gg} rigorously showed the necessity for a third Euler angle $\gamma$ in order to obtain a unique precessing frame which they dubbed the minimal-rotation frame.

The size of the parameter space for generic precessing binaries presents another  
formidable challenge for parameter estimation as the number of intrinsic parameters
increases from three (mass ratio and two spin magnitudes) 
for configurations where the spins are (anti)parallel to the orbital angular momentum,
which we refer to as either non-precessing or aligned-spin configurations,
to seven for binary black holes,
and even more in the case of binary neutron stars to additionally parametrize their tidal interactions.
As brute-force coverage of such a large space is computationally expensive,
approaches aimed at reducing the computational burden without compromising
waveform accuracy have emerged.
Of particular importance is Ref.~\cite{Schmidt:2012rh} which used an effective parametrization reducing the number of parameters 
to two in the QA frame by introducing 
an effective spin parameter\!\!
\footnote{To our knowledge, a similar parameter was first introduced in Ref.~\cite{Cutler:1994ys}, but not
for the same purpose.}\!\!
\;$\chi_\text{eff}$. 
The precessing waveform is then obtained by twisting the QA waveform with three Euler
angles as already described.
Ref.~\cite{Schmidt:2014iyl} took this approach further by packaging the four in-plane
(perpendicular to the Newtonian angular momentum) components of the binary's spin vectors into a single effective precession parameter, $\chi_p$, thereby reducing the dimensionality of the
parameter space to four.
On a parallel front, methods based on reduced-basis/order modelling were developed for generating fast, non-precessing waveforms
\cite{Field:2011mf, Field:2012if, Canizares:2013ywa, Purrer:2014fza, Canizares:2014fya, Purrer:2014fza, Purrer:2015tud, Bohe:2016gbl, Doctor:2017csx, Cotesta:2018fcv, Dietrich:2019kaq, Cotesta:2020qhw, Matas:2020wab}.
And finally, NR-``trained'' precessing waveform surrogates \cite{Blackman:2014maa, Blackman:2017dfb, Varma:2019csw, Williams:2019vub} have emerged
as the number of precessing NR simulations increased \cite{SXS:catalog}. 
Other approaches are also being developed such as ``the two-harmonic approximation'' \cite{Fairhurst:2019vut}.

In summary, there now exist several diverse precessing waveform approximants
of which the most prominent ones are \NRsurP{} \cite{Varma:2019csw}, 
$\mathtt{IMRPhenomPv3HM}$ \cite{Khan:2019kot}
(previously \texttt{PhenomPv2}), and \SEOBNR{} \cite{Ossokine:2020kjp}
(previously $\mathtt{SEOBNRv3}$).
These three approximants (along with the more recent \texttt{IMRPhenomXPHM}
\cite{Pratten:2020ceb}) have quickly become 
the preferred waveform models for parameter estimation by the LVC. 
However, they do not agree perfectly, 
which can lead to biases
as was illustrated, e.g., by Ref.~\cite{Williamson:2017evr} via an
\texttt{SEOBNRv3}-\texttt{IMRPhenomPv2} comparison,
highlighting what one should always keep in mind: 
waveform approximants are approximate
as the name implies so they can disagree,
therefore it is beneficial to have several approximants.

This paper is the first of a series that develops \TEOBResumP{}, a 
generic-spin approximant based on the Euler 
rotation of aligned-spin waveforms generated by \TEOBResumS~ \cite{Nagar:2018zoe}.
\TEOBResumS~is a state-of-the-art aligned-spin EOBNR model with enhanced 
spin-orbit, spin-spin, and tidal interactions \cite{Nagar:2018plt, Akcay:2018yyh}
that is very fast \cite{Nagar:2018gnk} and robustly produces inspiral-merger-ringdown 
waveforms for five additional modes beside the dominant (2,2) mode
\cite{Nagar:2019wds, Nagar:2020pcj}.
\TEOBResumS{} is very different in its design from
\SEOBNR{}, in particular in the spin sector \cite{Rettegno:2019tzh},
thus provides the only fully independent waveform model from the
approximants currently in use for GW analysis
(e.g., \texttt{PhenomPv3} uses fits of \texttt{SEOBNR} waveforms \cite{Ossokine:2020kjp}).
Our goal in this initial implementation of \TEOBResumP{} is to introduce minimal modifications to the existing \TEOBResumS{} infrastructure.
Therefore, we opt for an approach whereby we produce aligned, \emph{constant} spin waveforms
using \TEOBResumS{} then generate inspiral-merger precessing waveforms 
by twisting 
the non-precessing waveforms as is done in
the \texttt{IMRPhenomP}, \texttt{SEOBNR}, and \texttt{NRSur}{}
families.
We delegate the attachment of the ringdown portion of the precessing
waveforms to the next version of \TEOB.

This article is organized as follows.
We start by introducing the PN precession equations in Sec.~\ref{Sec:prec_phys}. 
In Sec.~\ref{Sec:Twist}, we present
details for the waveform twist operation. 
In Sec.~\ref{Sec:SpinTaylorT4_NR}, we compare $\TEOBResumP$ waveforms with the following waveform approximants: \NRsurP, \texttt{IMRPhenomPv3HM}, and \SEOBNR. We summarize our results in Sec.~\ref{Sec:end}.
We work in geometrized units setting $G=c=1$ from which one can recover the SI units
via $G M_\odot/c^3 \approx 4.925491\times 10^{-6}\,$sec, where $\Msun$ denotes a solar mass. 
We use bold font to denote Euclidean three-vectors with an overhat representing three-vectors of unit length. Overdots denote derivatives with respect to time.


\section{An Overview of Precessing Compact Binary Systems}\label{Sec:prec_phys}
Let us consider a compact binary system in a quasi-spherical inspiral 
with the subscript 1 labelling the primary and 2 labelling the secondary component.
Accordingly, the individual masses are denoted by $m_1$ and $ m_2$ with $m_1 \ge m_2$.
The total mass is defined as $M = m_1+m_2$. 
Let us also introduce the mass ratio $q\equiv m_2/m_1 \le 1$, the reduced mass $\mu\equiv M q/(1+q)^2$, 
and the symmetric mass ratio $\eta \equiv q/(1+q)^2$.
Note that in this article, we often set $M=1$, e.g., Eqs.~\eqref{eq:S1dot_NLO}-\eqref{eq:LNdot_NLO},
but sometimes restore solar-mass units ($M_\odot$) for $M$, cf.  Eqs.~\eqref{eq:S_perpMax}, \eqref{eq:Num_prec_cycles}.
We additionally endow the binary components with spins $\Sa, \Sb$, respectively, where $\mbf{S}_i\equiv m_i^2 \bm{\chi}_i$ with
$|\bm{\chi}_i|\le 1$ for $i=1,2$.

\subsection{Spin-orbit precession equations}\label{sec:prec_eq}

The Newtonian orbital angular momentum for the binary is given by 
$\LN= {\mu}\,\mathbf{r} \times \mathbf{v}$,
where $\mathbf{r}, \mathbf{v}$ are the relative separation and velocity vectors
of the binary in the usual center-of-mass frame.
Note that $\LN$ is different from its non-Newtonian counterpart
$\mbf{L}=\mathbf{r}\times \mathbf{p}$, where $\mbf{p}$ is the relative momentum.
This distinction, due to $\mu \mbf{v} \ne \mbf{p}$, is a consequence of the fully general relativistic 
MPD equations for the motion of a spinning test mass in curved spacetime.
From PN theory, one obtains $\mbf{L}= \LN + \Delta\mbf{L}_\text{1PN}+\ldots $ 
with correction terms, $\Delta\mbf{L}_{n\text{PN}}$, known up to 3.5PN (see, e.g., Eq.~(4.7) of Ref.~\cite{Bohe:2012mr}).
Note that, by definition, the Newtonian $\LN$ remains perpendicular to the orbital plane.

Let $\omega$ be the orbital frequency. Then, via Kepler's third law: 
$r\equiv |\mbf{r}|=\omega^{-2/3}$.
Accordingly, $\text{L}_\text{N} \equiv |\LN| = \mu r^2\omega=m_1 m_2/\omega^{1/3}= \eta/v$, where
we have introduced $v \equiv |\mbf{v}| = \omega^{1/3}$, i.e.,
the relative speed between the binary's components in the usual center-of-mass frame. 
Clearly, $v<1$ and furthermore, $v\ll1$ for most of the inspiral (recall, $v=v/c$ in restored units).
Note that each power of $v$ corresponds to a half PN order.
In this work, we use $v$ to track the orders in the precession equations.
Consequently, we reserve expressions such as 
next-to-leading order (NLO) to verbally track each power of $v$ beyond a given leading-order (LO) expression.

One can start with the general MPD equations of motion and obtain the PN expansions for the time evolution
of $\Sa, \Sb$. The details of this derivation can be found in, e.g., Secs.~ II, III of Ref.~\cite{Bohe:2015ana}, 
and Sec.~II of Ref.~\cite{Racine:2008kj}.
Up to NLO, i.e., 0.5PN, the orbital angular momentum and spin precession equations are given by \cite{Apostolatos:1994mx, Kidder:1995zr}
\begin{subequations}
\begin{align}
 \Sadot^\text{NLO} & =v^5\,\eta\left(2+\f{3}{2}q\right)\left(\Lhat\times \Sa\right) \label{eq:S1dot_NLO}\\
 &+\f{v^6}{2}\left\{\Sb-3 [(q\Sa+\Sb)\cdot\Lhat]\,\Lhat\right\}\times\Sa 
 , \nn\\
 \Sbdot^\text{NLO} &=v^5\,\eta\left(2+\f{3}{2q}\right)\left(\Lhat\times \Sb\right) \label{eq:S2dot_NLO}\\
 &+\f{v^6}{2}\left\{\Sa-3 [(\Sa+q^{-1}\Sb)\cdot\Lhat]\,\Lhat\right\}\times\Sb 
 ,\nn\\
 \Lhatdot^\text{NLO} &= -\f{v}{\eta} \left(\Sadot^\text{NLO} +\Sbdot^\text{NLO}\right) 
 \label{eq:LNdot_NLO}.
 \end{align}
\end{subequations}
Note that, as is usual in the literature, we present the orbit-averaged evolution equations. As such, our solutions to these equations do not capture the nutation
of $\LN$, but this is of no consequence for parameter estimation purposes
at the sensitivity of the advanced GW detectors \cite{Pan:2013rra}. 
For non-averaged versions, cf. App.~A of Ref.~\cite{Bohe:2015ana}.

The particular form of Eq.~\eqref{eq:LNdot_NLO} above is the result of total angular momentum conservation:
$\dot{\mbf{J}}=0$, where $ \mbf{J}=\mbf{L} +\S$ with $\S\equiv \Sa+\Sb$.
The forms of Eqs.~(\ref{eq:S1dot_NLO}-\ref{eq:LNdot_NLO}) have the added benefit that the evolution of the Newtonian orbital angular momentum can be written as a classical mechanical precession equation:
\be
\Lhatdot^\text{NLO} = \bm{\Omega}_\text{NLO}\times \Lhat ,\label{eq:LNhat_prec_eq} 
\ee
where $\bm{\Omega}_\text{NLO}$ can be extracted straightforwardly from Eqs.~(\ref{eq:S1dot_NLO} - 
\ref{eq:LNdot_NLO}).

The effect of radiation reaction is implicit in $v=v(t)$ in Eqs.~(\ref{eq:S1dot_NLO}-\ref{eq:LNdot_NLO}). 
For nonspinning systems, $\dot{v}=\dot{v}(v)$ is fully known as a PN series 
starting from $\sim v^9$ and going up to 3.5PN order $\sim v^{16}$.
For systems with spin, spin-orbit terms enter first at 1.5PN and spin-spin terms at 2PN.
Here, we employ the \texttt{TaylorT4} resummed form of $\dot{v}(v)$
\cite{Buonanno:2002fy, Buonanno:2009zt}
as adopted in the \STT{} approximant.
The series coefficients for $\dot{v}(v)$ can be found, e.g., in App.~A of Ref.~\cite{Chatziioannou:2013dza}.

For precessing binaries, there are three time scales of relevance: radiation-reaction timescale $T_\text{RR}$,
precession time scale $T_\text{pr}$, and orbital time scale $T_\text{orb}$.
Integrating $\dot{v}\sim v^9$ yields $T_\text{RR} \sim v^{-8}$. From $v= \omega^{1/3}$, 
we obtain $T_\text{orb} \sim v^{-3}$. Finally, the precession equation \eqref{eq:S1dot_NLO} gives $T_\text{pr}\sim|\Sa|/|\Sadot|\sim v^{-5}$.
Since $v \ll1$ mostly, we have the following separation of timescales:
\be
T_\text{orb} \ll T_\text{pr} \ll T_\text{RR} \label{eq:time_scales}.
\ee
Thanks to this separation of scales, we expect our hybrid approach,
which combines EOB dynamics with PN precession, to work well
as we show in Sec.~\ref{Sec:SpinTaylorT4_NR}. 

Recall that Eqs.~(\ref{eq:S1dot_NLO} - \ref{eq:LNdot_NLO}) are 0.5-PN (NLO) accurate.
Though this is the usual order in the literature, 
we employ versions of the precession ODEs that have been pushed to the limit 
of the current analytical PN knowledge, which we denote as N4LO (2\,PN) here.
As far we can tell these have never appeared in a journal article, but exist
in written form in several approximants such as \STT.
Defining $\delta m = m_1 - m_2$ in natural units [e.g., $m_1=1/(1+q)$],
the N4LO spin-orbit precession ODEs read
%
\begin{widetext}
\begin{subequations}
\begin{align}
\Sadot^\text{N4LO} &= \ \Sadot^\text{NNLO}+ v^9\left[\f{27}{32}+\f{3\eta}{16}-\f{105\eta^2}{32} 
-\f{\eta^3}{48}+\delta m\left(-\f{27}{32}+\f{39\eta}{8}-\f{5\eta^2}{32} \right) \right](\Lhat\times\Sa), \label{eq:S1dot_N4LO}\\
\Sbdot^\text{N4LO} &=  \ \Sbdot^\text{NNLO}+ v^9\left[\f{27}{32}+\f{3\eta}{16}-\f{105\eta^2}{32} 
-\f{\eta^3}{48}-\delta m\left(-\f{27}{32}+\f{39\eta}{8}-\f{5\eta^2}{32} \right) \right](\Lhat\times\Sb), \label{eq:S2dot_N4LO}\\
\Lhatdot^\text{N4LO} & =\text{L}^{-1}_{2\text{PN}} \left[  \f{v}{\eta} \left(-\Sadot^\text{N4LO}-\Sbdot^\text{N4LO} \right)
- v^3\left(c_{S1} \Sadot^\text{NNLO}+c_{S2}\Sbdot^\text{NNLO}\right)\right.\label{eq:LNhatdot_N4LO}\\
 &\qquad \qquad\left.- v^3 \left\{c_{S1L}\left(\left[\f{-v}{\eta}(\Sadot^\text{NLO}+\Sbdot^\text{NLO})(\Lhat\cdot\Sa) \right]
+\Lhat\left[ -\f{v}{\eta}\Sbdot^\text{NLO}\cdot\Sa+\Lhat \cdot \Sadot^\text{NNLO}\right]\right)+(1\leftrightarrow 2) \right\}\right] \nn,
\end{align}
\end{subequations}
\end{widetext}
%
where 
\begin{subequations}
\begin{align}
\Sadot^\text{NNLO} &= \ \Sadot^\text{NLO}\nn \\
+& v^7\left[\f{9}{16}+\f{5\eta}{4}-\f{\eta^2}{24}
+\delta m \left(-\f{9}{16}+\f{5\eta}{8}\right) \right](\Lhat\times\Sa) \label{eq:S1dot_NNLO},\\
\text{L}_{2\text{PN}} &=1 + v^2\left(\f{3}{2}+\f{\eta}{6}\right)
+ v^4\left( \f{27}{8}-\f{19\eta}{8}+\f{\eta^2}{24}\right) \label{eq:L_2PN},
\end{align}
\end{subequations}
\begin{subequations}
\begin{align}
c_{S1}&=-\f{1}{4}\left(3+\f{1}{m_1}\right)\label{eq:c_S1},\\
c_{S1L}&=-\f{1}{12}\left(1+\f{27}{m_1}\right) \label{eq:c_S1L}.
\end{align}
\end{subequations}
$\Sbdot^\text{NNLO}, c_{S2},c_{S2L} $ can be obtained via the $(1\leftrightarrow 2)$ exchange.

Note that from NNLO on, one no longer has a standard precession equation for $\Lhatdot$ of the form of
Eq.~\eqref{eq:LNhat_prec_eq}. In fact, as can be seen from Eq.~\eqref{eq:LNhatdot_N4LO}, 
$\Lhatdot$ has components both perpendicular and \emph{parallel} to $\Lhat$. 
Therefore, we define \cite{Sturani_note}

\be
\LhatdotNew^\text{N4LO}  \equiv \ \Lhatdot^\text{N4LO} - (\Lhat\cdot \Lhatdot^\text{N4LO})\Lhat \label{eq:Lperpdot_N4LO},
\ee
which then satisfies
\be
\LhatdotNew^\text{N4LO} = \bm{\Omega}_L^\text{N4LO} \times \Lhat^\text{N4LO} \label{eq:LNhat_prec_eq2} .
\ee
We use the solutions of Eq.~\eqref{eq:Lperpdot_N4LO}\!\!
\footnote{These expressions match their \texttt{spinOrd} = 7 counterparts
as given in the \STT{} approximant.}, to compute $\Lhat(t)$,
but we have also used directly the solutions to Eq.~\eqref{eq:LNhatdot_N4LO} 
and found relative differences in the components of $\Lhat$ of $\lesssim 10^{-4}$.
We present the derivational details of these N4LO expressions in App.~\ref{sec:AppA}.

There are indications that the PN precession equations converge with increasing
PN order despite missing higher-order information \cite{Ossokine:2015vda}.
Indeed, we have found it slightly more beneficial to work with the N4LO precession equations
rather than the NLO versions. 
We illustrate this is in App.~\ref{sec:AppB},
where we show that the N4LO-Euler-angle twisted \TEOB{}
agrees better with both \NRsurP{} and \SEOBNR{}
than its NLO counterpart. 
This agreement is demonstrated specifically in terms of waveform strain mismatches
which we introduce in Sec.~\ref{Sec:SpinTaylorT4_NR}.
The NLO-N4LO disagreement is more severe for systems with more mass asymmetry,
i.e., smaller values of $q$, which we show in terms of Euler angles
in Fig.~\ref{fig:Euler_angles} in App.~\ref{sec:AppA}.
As the figure exhibits, there is considerable Euler-angle dephasing
between NLO, NNLO, and N4LO solutions for small $q$, but no such dephasing
between N3LO and N4LO, which we somewhat expect since their difference
is at 2\,PN. We discuss the various ODE orders further in App.~\ref{sec:AppA}
We should add that instantaneous corrections
to the orbit-averaged expressions start entering at 
N3LO \cite{Bohe:2015ana} which we do not take into account here,

As already mentioned, it is useful to package the six spin degrees of freedom into a space of lower dimensions.
This is usually done by considering the projections of $\Sa,\Sb$ parallel and orthogonal to $\Lhat(t)$,
resulting in two commonly employed scalar quantities.
The parallel scalar is \cite{Ajith:2011ec, Damour:2001tu, Racine:2008qv}
\be
\chi_\text{eff}= M^{-2}\left[(1+q)\Sa\cdot\Lhat+(1+q^{-1})\Sb\cdot \Lhat\right] \label{eq:chi_eff}
\ee
which is a conserved quantity of the orbit-averaged precession equations over the precession timescale \cite{Racine:2008qv}.
The orthogonal parameter is $\chi_p$ of Ref.~\cite{Schmidt:2014iyl} defined as
\!\!\footnote{Note that the factor in front of $\max\{\ldots\}$ may differ depending on the convention
that assigns either $m_1$ or $m_2$ as the primary binary component.}
\be
\chi_p \equiv \f{m_1^{-2}}{\left(2+3q/2\right)}\max\left\{\!\left(2+\f{3q}{2}\right)\!|\mbf{S}_{1,\perp}|,
 \left(2+\f{3}{2q}\right)\!|\mbf{S}_{2,\perp}|\right\} \label{eq:chi_p},
\ee
where $\mbf{S}_{1,\perp},\mbf{S}_{2,\perp}$ denote the components of $\Sa(t),\Sb(t)$ perpendicular to
$\Lhat$, respectively. Both $\chi_\text{eff}$ and $\chi_p$ are commonly used in the LVC analysis of GW events \cite{LIGOScientific:2018mvr}.

We now introduce a new orthogonal parameter
%
\begin{align}
 \chi_{\perp,\max} &\equiv M^{-2}\max \left|\, \mbf{S}_{1,\perp}+\mbf{S}_{2,\perp}\right|, \label{eq:S_perpMax}
\end{align}
%
where we take the maximum value of the norm over the entire time evolution.
$\Sperp$ seems to encode the strength of precession
as we show in Secs.~\ref{sec:twist_NRsur}-\ref{sec:twist_SEOBNR}.
Note that $\Sperp$ is bounded above by $(1+q^2)/(1+q)^2$ which yields 
0.5 for $q=1$ and 1 in the test-mass limit.

\subsection{Reference frames}\label{sec:Frames}
When considering precessing systems, there are two special frames of reference which have their respective $z$-axes aligned with 
$\Livec\equiv {\mathbf{L}}_\text{N}(t_0)$ and ${\mathbf{L}}_\text{N}(t)$,
where $t_0$ is some arbitrary time at the initial configuration of each binary. It is common in the waveform
community to set $t=0$ to coincide with the peak of the non-precessing (2,2)
mode, which then gives us $t_0<0$. In what follows, we assume a constant
shift in $t$ such that the initial time is given by $t_0=0$ with the peak time positive as in done in \TEOBResumS{} waveforms. 
Therefore, we write $\Livec=\LN(0)$ and similarly for all 
other relevant quantities.
We refer to the $\LN(0)$ and $\LN(t)$ frames as the $\Livec$ frame, and the co-precessing frame, respectively. 
Clearly, the $\Livec$ frame is inertial whereas the co-precessing frame is not.
One can additionally introduce a second inertial frame, $\Jivec$, 
where one aligns the $z$-axis
with $\hat{\mathbf{J}}$ (Newtonian, 1PN or 2PN) either at the initial time or
at the peak of the orbital frequency. For reasons that we explain in 
Sec.~\ref{sec:qualitative_disc_prec},
we choose $\Jivec\equiv {\mathbf{J}}_\text{2PN}(0)$ obtained from
the N4LO solutions for $\LN,\Sa,$ and $\Sb$.
Note that our $\Jivec$ is different than the one introduced in
Ref.~\cite{Apostolatos:1994mx}, which is given by
$\hat{\mathbf{J}}_\text{N} - \epsilon\, \hat{\mathbf{J}}_\text{N} \times \!\hat{\,\mathbf{L}}_\text{N}$, 
where $\epsilon \ll 1$ and the ODEs are truncated at NLO with the spin-spin
term additionally turned off so that $(d/dt) |\Sa+\Sb|=0$ \cite{Apostolatos:1994mx}.

\begin{figure}[t]
    \centering
    \includegraphics[width=0.5\textwidth]{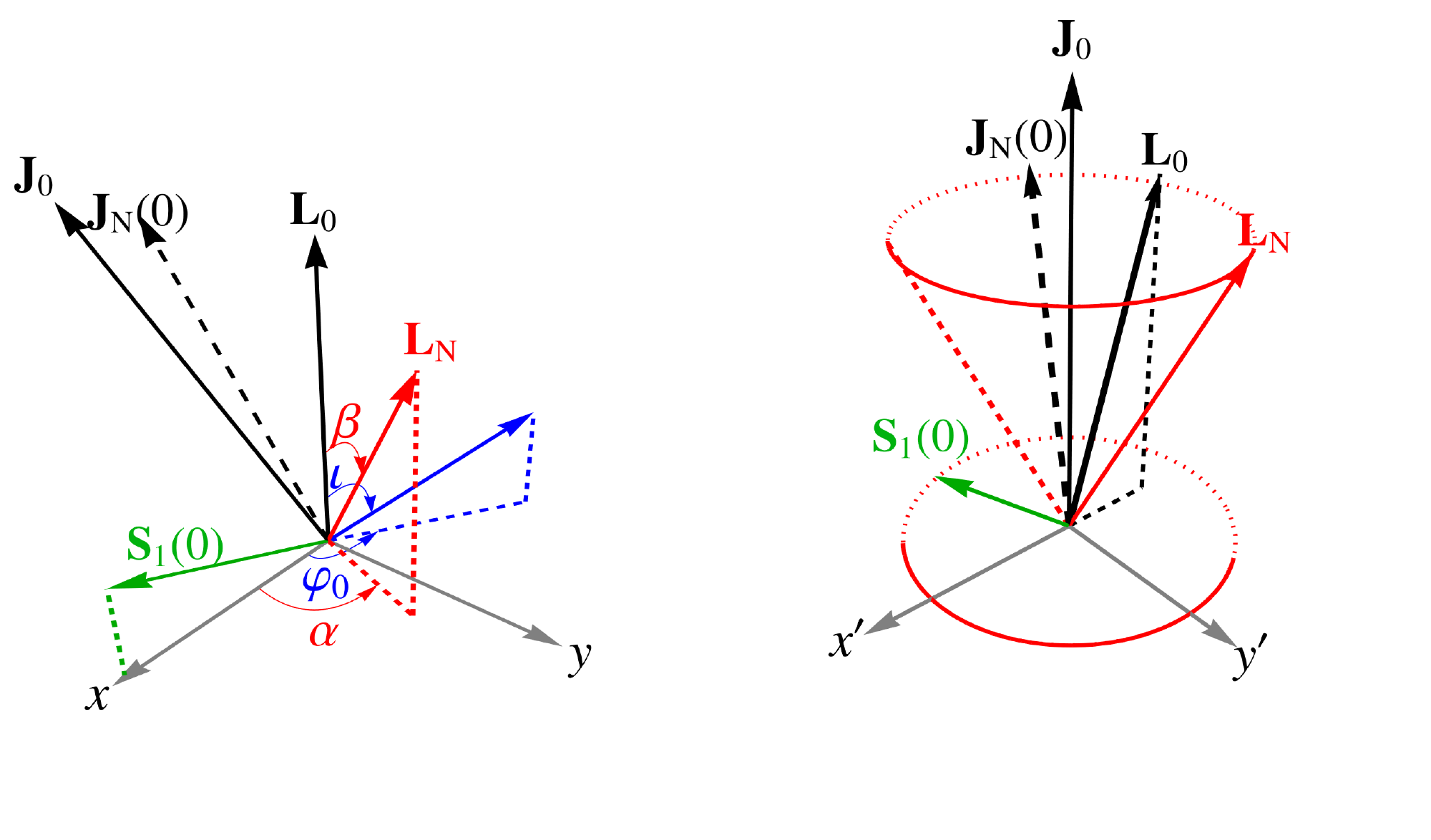} 
\caption{The inertial $\Livec$ and $\Jivec$ frames whose $z$-axes are parallel
to $\LN(0) $ and $\J_\text{2PN}(0) $, respectively.
We choose the $x$-axis of the $\Livec$ frame 
such that the initial spin of the primary component, $\Sa(0)$ lays in the $x$-$z$ plane.
In this frame, we denote the spherical angles of $\LN\equiv \LN(t)$ (red arrow) by $\alpha$
and $\beta$. In the $\Jivec$ frame, it is easier to discern the precession of $\LN$
as it approximately traces out a cone per precession cycle (only approximately
because $|\LN|$ decreases due to radiation reaction, see Sec.~\ref{sec:qualitative_disc_prec}).
We show such a cone in the right-hand panel along with its projection onto the plane
perpendicular to $\Jivec$. 
We also show $\JN(0)$ as the dashed arrow, which is slightly different than
our $\Jivec$ which we explain in Sec.~\ref{sec:qualitative_disc_prec}.
The blue arrow with the polar angles $(\iota,\varphi_0)$ represents the line of sight to the detector.
}
\label{fig:Frames}
\end{figure}

The $\Livec$ frame is our preferred frame here
as it is the most straightforward frame for solving the precession ODEs \eqref{eq:S1dot_NLO} - \eqref{eq:LNdot_NLO}
even though the precession-induced amplitude modulations are more pronounced in this frame.
Accordingly, we label the azimuthal and the polar angles of $\LN(t)$ with respect to $\Livec$ 
by $\alpha$ and  $\beta$ as shown in Fig.~\ref{fig:Frames}.

Naturally, we must pick an $x$-axis in the $\Livec$ frame, with respect to which we measure $\alpha(t)$.
Here, as in Ref.~\cite{Buonanno:2002fy}, we impose the condition that $\Sa(0)$ is in the $x$-$z$ plane
as shown in Fig.~\ref{fig:Frames},
which yields
$\hat{\bm{x}} =\mbf{S}^0_{1,\perp}/|\mbf{S}^0_{1,\perp}|$, where $\mbf{S}^0_{1,\perp}\equiv \Sa(0) -(\Sa(0)\cdot \LOhat)\LOhat$
and $\LOhat \equiv \Livec/|\Livec|$.
We can therefore fully specify $\Sa(0)$ via the parameters $\{q,\chi_1,\theta_1\}$ 
where $\theta_1 = \cos^{-1}(\Sa(0)\cdot \!\hat{\,\mbf{L}}_0/S_1)$,
$S_1\equiv|\Sa|=\chi_1 (1+q)^{-2}$
with $m_1=M/(1+q)$ setting $M=1$ and assuming $q\le 1$.
Similarly, $\Sb(0)$ is specified by $\{q,\chi_2,\theta_2,\phi_2\}$ where 
$\theta_2 = \cos^{-1}(\Sb(0)\cdot\LOhat/S_2)$, $S_2\equiv|\Sb|=\chi_2 \,q^2(1+q)^{-2}$ and $\phi_2$ is the azimuthal angle with respect to $\hat{\bm{x}}$ defined above.
With our axes defined, it is straightforward to obtain
\begin{subequations}
\begin{align}
\alpha &= \tan^{-1} \left(\f{\mbf{L}_{\text{N}}\cdot \hat{\bm{y}}}{\mbf{L}_{\text{N}}\cdot\hat{\bm{x}}} \right)\label{eq:alpha},\\
\beta & = \cos^{-1} \left(\!\hat{\,\mathbf{L}}_{\text{N}}\cdot \LOhat\right)\label{eq:beta}.
\end{align}
\end{subequations}
The third angle, as introduced by Ref.~\cite{Boyle:2011gg}, is given by the solution to
$\dot{\gamma}=\dot{\alpha}\cos\beta$, where we chose to keep the right-hand side positive 
to have $\gamma(t)$ monotonically increasing like $\alpha(t)$.

%
%

Note that, for the purposes of data analysis and parameter estimation,
we must restore $M$ to its physical units which we denote by $M_\text{tot}(\Msun)$.
This is because the detection band of the GW interferometers is roughly between
$20$ and $2000\,$Hz and the heavier binary systems merge at lower frequencies.
Therefore, we parametrize our precessing binary inspirals using the 
following finalized set consisting of eight parameters
\be
\{ \fgwi(\text{Hz}), M_\text{tot}(\Msun),q,\chi_1,\chi_2, \theta_1, \theta_2,\phi_2\} \label{eq:parameters},
\ee
where $\fgwi$ is the initial $(2,2)$-mode GW frequency marking the starting point of each inspiral.

\subsection{Effects of precession}\label{sec:qualitative_disc_prec}
Spin-orbit precession occurs when the spins are not (anti)parallel to the orbital angular momentum.
The main effect is a slow precession of $\LN$ about an axis that is
roughly aligned with $\J(0)$, but the true fixed axis depends on the
order at which the precession ODEs are truncated, and the use of the appropriate solutions to those ODEs.
For example, in Ref.~\cite{Apostolatos:1994mx}, this axis is given by
$\Jivec\equiv \hat{\mathbf{J}}_\text{N} - \epsilon\, \hat{\mathbf{J}}_\text{N}\times \!\hat{\,\mathbf{L}}_\text{N}$ 
obtained from the NLO solutions while neglecting the spin-spin coupling.
This $\Jivec$ then indeed remains fixed.
However, we neither truncate the ODEs at NLO, nor neglect the spin-spin coupling,
therefore we choose not to use this choice for $\Jivec$.
As we solve the N4LO (2PN) equations here, which are obtained by imposing 
$\dot{\mbf{J}}=\!\dot{\,\mbf{L}}+\dot{\mbf{S}}_1+\dot{\mbf{S}}_2=0$
with $\mbf{L}$ up to 3.5PN decaying under radiation reaction (see App.~\ref{sec:AppA}), 
we have at best an approximate conservation
of $\J_\text{2PN}$. Therefore, we set $\Jivec=\J_\text{2PN}(0)$.

\begin{figure}[t]
\centering   
   \includegraphics[width=.49\textwidth]{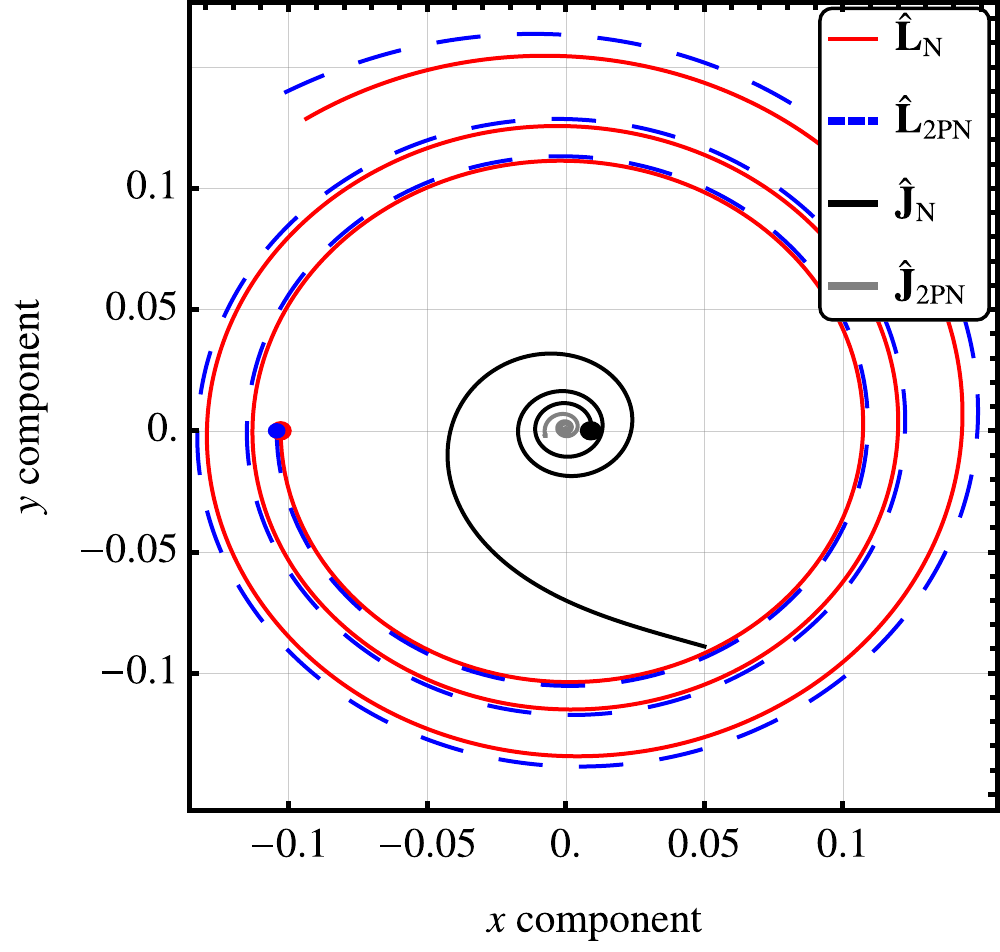}
\caption{\label{fig:LxLy_tracks} 
Tracks of the $x,y$ components of the Newtonian orbital ($\Lhat(t)$, red) 
and total angular [unit] momenta
($\Jhat(t)$, black) 
in the plane orthogonal to $\Jivec=\J_\text{2PN}(0)$
for a binary with $M_\text{tot}=30 M_\odot, q= 1, \chi_1= \chi_2=0.7, \theta_1= \theta_2=90^\circ,$ and $\phi_2=135^\circ$ starting from the GW frequency of 20\,Hz.  
We additionally show the components of the 2PN-corrected orbital angular momentum, $\!\hat{\,\mbf{L}}_\text{2PN}$ 
(dashed blue), and the corresponding total angular momentum,
$\hat{\mbf{J}}_\text{2PN}$ (gray). The dots mark the starting positions for each vector.
As described in the text, the various angular momenta spiral outward
around the fixed axis $\Jivec$, but
$\hat{\mbf{J}}_\text{2PN}$ outspirals much less than $\Jhat$
consistent with our use of solutions to the 2PN spin precession equations.
}
\end{figure}

The decrease of $\LN$ under radiation reaction
results in a precession cone whose opening angle increases in time
as illustrated in Ref.~\cite{Apostolatos:1994mx}.
Thus, the projection of $\Lhat$ orthogonal to $\Jivec$
shows circularly outspiraling tracks as in Fig.~\ref{fig:LxLy_tracks}.
Furthermore, $\Jhat$ also precesses around, in fact, 
out-spirals around $\Jivec$, 
which we also exhibit in Fig.~\ref{fig:LxLy_tracks}.
This spiralling behavior persists for PN-corrected $\L$ and $\J$,
albeit with smaller precession cone opening angles for $\J$,
as we show for $\hat{\mbf{J}}_{2\text{PN}}$ in the figure.
For three-dimensional versions of these, see Apostolatos \etal~\cite{Apostolatos:1994mx} 
which still remains the most illustrative resource for understanding the qualitative behavior of precessing systems.
Ref.~\cite{Apostolatos:1994mx} also provides a useful expression for the number of precession cycles when the masses are small and initial separation is large, i.e.,  $ |\LN| \gg |\Sa+\Sb|$,
\be 
N_\alpha\equiv \f{\alpha}{2\pi}\approx 11\left(1+\f{3 m_1}{4m_2}\right) \f{10 M_\odot}{\Mtot} \f{10\,\text{Hz}}{f} \label{eq:Num_prec_cycles},
\ee
where, recall $M_\text{tot}$ is $M$ in solar masses.
%

\begin{figure}[t]
\centering   
   \includegraphics[width=0.49\textwidth]{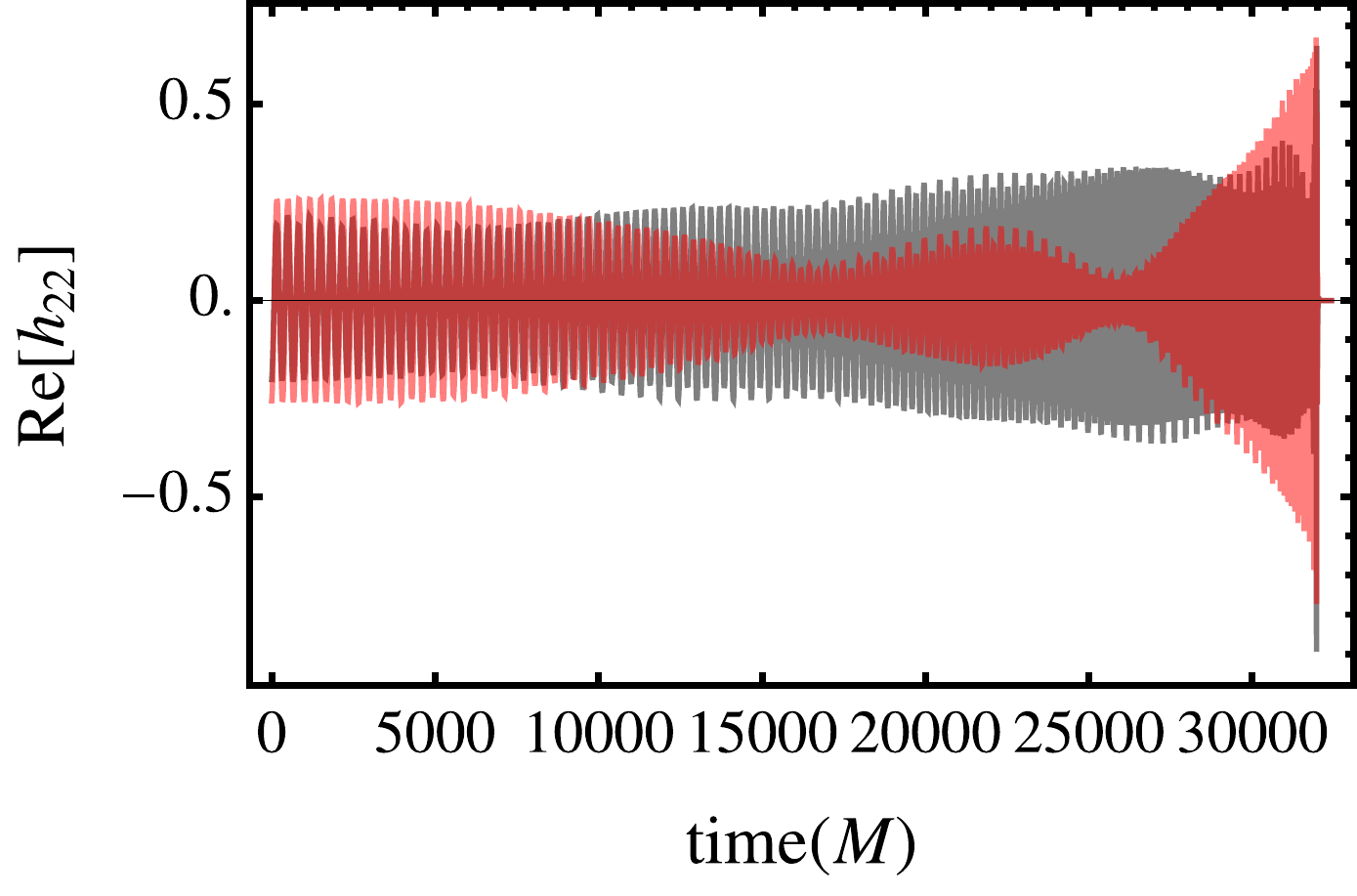}
\caption{\label{fig:Two_prec_waves}
Precessing $(2,2)$ modes as viewed by observers whose line of sight is parallel to $\Livec$ (red)
and to $\Jivec$ 
(gray) for a binary system with
$M_\text{tot}=30 M_\odot, q= 1/5, \chi_1=\chi_2=0.7, \theta_1=\theta_2=\phi_2=135^\circ$
starting from 20\,Hz. 
As discussed in the text, the $\Livec$-frame observers see much more pronounced amplitude modulations
than their $ \Jivec$-frame counterparts.
}
\end{figure}

The precession of $\LN$ induces amplitude modulations in the waveform and modifies the phase.
The modulations depend strongly on the orientation of the orbit with respect to an observer's
line of sight. This is illustrated in Fig~\ref{fig:Two_prec_waves}, where the gray curve is the 
precessing $(2,2)$ mode as seen by an observer lined up with $\Jivec$ 
who receives less modulated GWs
because $\LN(t)$ tracks a circularly inspiralling path as 
depicted in Fig.~\ref{fig:LxLy_tracks} whereas the $\Livec$-frame
observer sees emissions over an elliptically inspiralling track, 
hence resulting in larger amplitude variations.
This means that the reference frame in which the incoming GWs are received (e.g., detector frame)
plays a significant role in GW detection as using non-precessing waveform template banks to match-filter the signal
can lead to a significant fraction of precessing signals being missed or dismissed as glitches \cite{Schmidt:2012rh}.

Thus far, we have talked about simple precession dubbed so because both $\L$ and $\S$ precess around $\J$.
However, when $\L+\S \approx 0$, a phenomenon known as 
transitional precession occurs in which $\J$ ``tumbles'' until radiation
reaction decays $\L$ enough to take the system away from the $\L+\S \approx 0$ configuration \cite{Apostolatos:1994mx}.
Since transitional precession requires careful fine-tuning of the parameters, it is expected to be a rare
phenomenon \cite{Apostolatos:1994mx, Buonanno:2002fy} so
we do not consider it here.


\section{Twisting non-precessing waveforms}\label{Sec:Twist}
Having conceptually introduced the twist operation, we next provide mathematical details. Our discussion here is mostly based on 
Refs.~\cite{Boyle:2011gg, Schmidt:2012rh, Babak:2016tgq}.
Let us recall that $\alpha(t)$ and $\beta(t)$ are the azimuthal and polar angles of $\LN(t)$ with respect
to $\Livec=\LN(0)$ and the third angle $\gamma(t)$ is obtained from $\dot{\gamma}=\dot{\alpha}\cos\beta$.
The set $\{\alpha(t),\beta(t),\gamma(t)\}$ is all we need when transforming between the 
$\LN(t)$ and $\Livec$ frames.
Specifically, when going from our inertial $\Livec$ frame to the $\LN(t)$-frame, we ``forward''-Euler rotate
using $R(\alpha,\beta,\gamma)\equiv R_z(\gamma)R_y(\beta) R_z(\alpha)$
where $R_j(\zeta_k)$ represent rotations by the angles $\zeta_k$ with respect to the $j$ axis\footnote{In this article, 
we use the $z$-$y'$-$z''$ convention for Euler rotations as is standard in the relevant literature \cite{Schmidt:2010it, Boyle:2011gg, Babak:2016tgq}. 
}. In the following, we omit displaying the explicit time dependence of these angles and various other
time-dependent quantities, e.g., $\LN(t)$, which we restore when necessary.

Under the forward Euler rotation above, 
the gravitational-wave modes transform as follows
\be
h_{\ell m} = \sum_{m'=-\ell}^\ell  h_{\ell m'} \, D^{(\ell)}_{m',m}(\alpha,\beta,\gamma)\, \label{eq:hlm_transformation},
\ee
where $ D^{(\ell)}_{m',m}$ are Wigner's D matrices which can be related to 
spin-weighted spherical harmonics via \cite{Goldberg:1966uu}
\be
{}_sY^{\ell m}(\theta,\phi) = (-1)^m \sqrt{\f{2\ell+1}{4\pi}} D^{(\ell)}_{-m,s}(\phi,\theta,0). \label{eq:sYlm_WigD_relation}
\ee
Note that different versions of this equation exist in the literature due to conventions of Wigner D
matrices.
Here, we employ the definition introduced in Ref.~\cite{Sakurai:1167961}
\be
D^{(\ell)}_{m',m}(\alpha,\beta,\gamma)=e^{-i m'\alpha} e^{-i m \gamma}d^\ell_{m',m}(\beta), \label{eq:WigD_explicit}
\ee
where $d^\ell_{m',m}(\beta)\in \mathbb{R}$ are the ``little'' D matrices given by
\begin{align}
d^\ell_{m',m}(\beta)=&\sum_{k_i}^{k_f}
(-1)^{k-m+m'}\nn \\
&\times\f{\sqrt{(\el+m)!(\el-m)!(\el+m')!(\el-m')!}}{k!(\el+m-k)!(\el-k-m')!(k-m+m')!}\nn \\
	      & \times \left[\cos\f{\beta}{2} \right]^{2\el-2k+m-m'} \left[\sin\f{\beta}{2} \right]^{2k-m+m'}
	      \label{eq:Wigd_explicit},
\end{align}
where $k_i=\min(0,m-m')$ and $k_f=\max(\el+m, \el-m')$.

As explained in Sec.~\ref{Sec:introduction}, the key idea
is to ``unwrap'' or twist aligned-spin waveforms generated in the $\LN$ frame using Euler rotations. 
In order to transform from $\LN$ to $\Livec$ frame,
we ``backward'' Euler-rotate via the inverse rotation matrices: $R^{-1}=R(-\gamma,-\beta,-\alpha)$.
Therefore, to twist we invert Eq.~\eqref{eq:hlm_transformation} \cite{Schmidt:2010it, Schmidt:2012rh}
\be
h^\text{T}_{\ell m} = \sum_{m'=-\ell}^\ell  h^\text{NP}_{\ell m'} \, D^{(\ell)\ast}_{m',m}(-\gamma,-\beta,-\alpha)\,,
\label{eq:hlm_Twist1}
\ee
where we introduced the superscripts T and NP to denote the twisted and the non-precessing waveforms, respectively.
Using the standard identity $D^{(\ell)\ast}_{m',m}= (-1)^{m'-m}D^{(\ell)}_{-m',-m}$ which translates to
$ (-1)^{m'-m}d^{\ell}_{-m',-m}(-\beta)= d^{\ell}_{m',m}(-\beta)$ in Eq.~\eqref{eq:WigD_explicit},
we obtain \cite{Schmidt:2010it, Schmidt:2012rh, Hannam:2013oca, Khan:2018fmp}
\be
h_{\ell m}^\text{T}(t) = e^{-i m \alpha(t)}\sum_{m'=-l}^l e^{i m'\gamma(t)} d^\ell_{m',m}(-\beta(t))\, h_{l m'}^\text{NP}(t)\label{eq:hlm_Twist2} \, ,
\ee
where we restored the time dependences.
\begin{widetext}

\begin{figure}[t]
\hspace{-0.9cm}
   \includegraphics[width=1.0\textwidth]{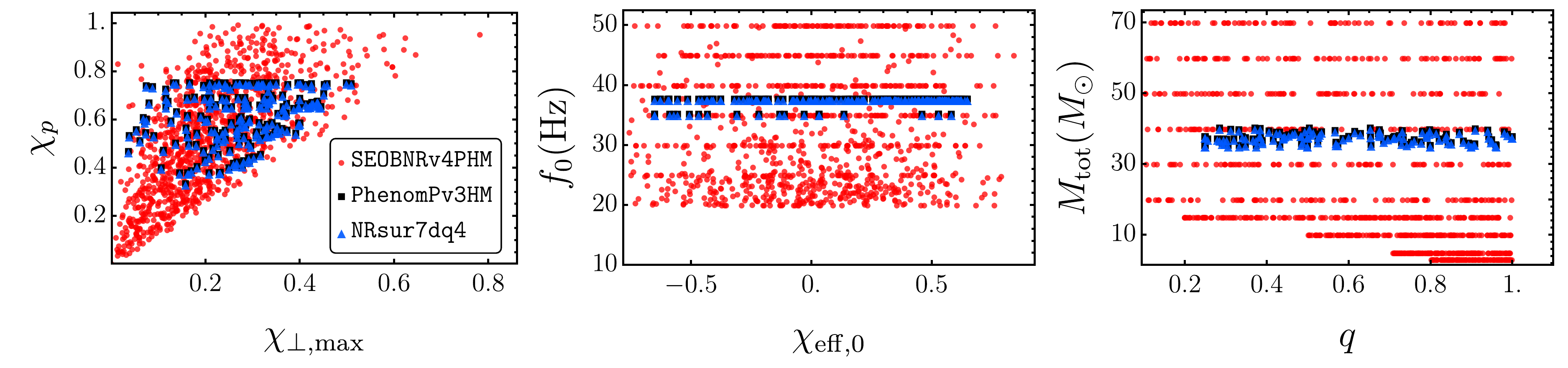}
\caption{\label{fig:parameter_space} 
Our coverage of the eight-dimensional parameter space of precessing compact
binary inspirals used in assessing the faithfulness of \TEOBResumP.
The parameters $\{\fgwi, q, \Mtot\} $ span the space of nonspinning
binaries, which is complemented by the set $\{\chi_{\text{eff},0},\chi_p\}$ 
or $\{\chi_{\text{eff},0},\Sperp\}$ that projects the five spin degrees of freedom
in $\Sa(0),\Sb(0)$ [by design $\mbf{S}_{1y}(0)=0$] to only two
via Eqs.~\eqref{eq:chi_eff}, \eqref{eq:chi_p}, and \eqref{eq:S_perpMax}.
For our assessment, we considered \nNR{} precessing binaries for the comparisons 
with \NRsurP{} (blue triangles), \PHM{} (black squares),
and {\nTot} binaries for comparisons with \SEOBNR{} (red dots).
Note that some parameters are duplicate within the $\{\fgwi,q,\Mtot\}$
subset, hence there are fewer points in the middle and right panels than the
rough total of \nTot.
}
\end{figure}
\end{widetext}

Note that the literature is replete with slightly different 
versions of Eq.~\eqref{eq:hlm_Twist2} depending on:
(i) Euler rotation conventions, (ii) Wigner D and spherical harmonic conventions, and (iii) the sign of
the right-hand-side for the $\dot{\gamma}$ equation. Our definitions and conventions agree with Ref.~\cite{Babak:2016tgq} (modulo the sign of $\gamma$) and our practical expression \eqref{eq:hlm_Twist2} agrees with Ref.~\cite{Khan:2018fmp} which interestingly disagrees with its updated version in Ref.~\cite{Khan:2019kot}, but then agrees with a recent
version used in \texttt{IMRPhenomXPHM} \cite{Pratten:2020ceb}.
We tested the performance of the alternate expression of Ref.~\cite{Khan:2019kot}
against ours in terms the $\el=2$ detector strain mismatches of \TEOB{} with \SEOBNR{} and \NRsurP{}. 
We found that the expression for the twist given by Eq.~\eqref{eq:hlm_Twist2}
performed better in the sense that it produced smaller mismatches. 
We delegate the details of this comparison to App.~\ref{sec:AppB}.

In principle, one can also twist the non-precessing waveforms using the angles of PN-corrected $\L(t)$
with respect to $\L(0)$. 
Ref.~\cite{Pan:2013rra} showed that the resulting differences in the twisted waveforms as compared 
with precessing NR waveforms are marginal, therefore we use only $\LN(t)$ with respect to
$\LN(0)$ for \TEOB.

We now have all the individual ingredients necessary to generate the precessing \TEOB{} waveforms.
The procedure for this operation is as follows:
\begin{enumerate}
\item Specify the initial parameters listed in Eq.~\eqref{eq:parameters}.
 \item Generate aligned-spin (non-precessing) $\ell=2$ waveform modes using \TEOBResumS{} 
 via the set of parameters $\{\fgwi,\Mtot, q,\chi_1,\chi_2\}$.
 \item Solve the orbit-averaged spin precession ODEs (\ref{eq:S1dot_N4LO}-\ref{eq:LNhatdot_N4LO}) using \STT{} resummed radiation reaction for $\dot{v}$.
 \item Retrieve the spherical angles $\{\alpha(t),\beta(t)\}$ from the components of $\LN(t)$ in the $\LN(0)$
 frame and subsequently obtain $\gamma(t)$ by solving $\dot{\gamma}=\dot{\alpha}\cos\beta$.
 \item Construct the precessing $\ell=2$ \TEOBResumP{} modes via the twist formula \eqref{eq:hlm_Twist2}.
\end{enumerate}

Let us conclude this section with three remarks: 
\begin{inparaenum}[(i)]
\item We can generate twisted waveforms in the $\JN(0)$ frame
as well as the $\LN(0)$ frame, but this is slower because
the solutions to the ODEs, which are solved in the $\LN(0)$ frame, 
must be Euler-rotated to the $\JN(0)$ frame at each time step.
Therefore, for convenience we compare in the $\LN(0)$ frame, 
but in principle we can straightforwardly rotate to the $\JN(0)$ frame.
\item It is possible to extend the above scheme
by coupling the precession ODEs to \TEOBResumS{} dynamics, i.e.,
by setting $\chi_i=\Lhat(t)\cdot\mbf{S}_{i}(t)/m_i^2$ ($i=1,2$)
at each time step of the aligned-spin EOB dynamics, 
where $\Lhat(t),\mbf{S}_i(t)$ are obtained from the N4LO precession dynamics.
This is similar to what is done in the precessing \texttt{SEOBNRv3,v4} 
approximants, where aligned-spin modes with time-\emph{varying} $\chi_i$ 
are twisted \cite{Pan:2013rra, Babak:2016tgq, Ossokine:2020kjp}.
\item For this initial version of \TEOBResumP, we truncate our precessing waveforms before the  transition to ringdown.
Attaching the ringdown portion to the inspiral-plunge-merger (IM) part of the precessing waveforms is quite a subtle procedure, especially in the 
time domain.
For example, in \texttt{SEOBNRv3P} \cite{Babak:2016tgq}, the ringdown waveforms are computed
in the $\mbf{J}(t_\text{match})$ frame, where $t_\text{match}$ approximates
the merger time. Then, the ringdown waveform is attached to the precessing IM portions obtained by twisting the co-precessing modes to the 
$\mbf{J}(t_\text{match})$ frame \cite{Babak:2016tgq}.
In the upgraded version, \SEOBNR{} \cite{Ossokine:2020kjp}, the ringdown is attached in the co-precessing
frame, which is less complicated to implement and less prone to numerical instabilities, thus is more appealing to us as a ringdown implementation.
There is also the question of how far one can push the PN ODEs.
The time domain \texttt{IMRPhenomTP} \cite{Estelles:2020osj} approximant provides a prescription for extending $\alpha(t), \beta(t)$ into the
ringdown regime (see Ref.~\cite{Marsat:2018oam} for details of this prescription)
and also uses the same implementation for $\alpha(t)$ as \SEOBNR.
However, Ref.~\cite{Estelles:2020osj} remarks that this is a ``simple implementation'' that will be improved.
In short, the ringdown attachment requires extreme care and detailed
testing, that is why we leave it for the next version of \TEOBResumP{}.
\end{inparaenum}


\section{Assessing the Twist: Comparisons with \NRsurP, \PHM{} and  \SEOBNR{} }\label{Sec:SpinTaylorT4_NR}%
To assess the faithfulness of \TEOBResumP{} (henceforth \TEO), 
we compared the twisted \TEOBResumS{} waveforms against precessing waveforms
generated by the following three approximants: \NRsurP, \PHM, and \SEOBNR{}
(henceforth, \NR, \Phn, \SEOB).
We first considered a set of \nNR{} precessing binaries
consisting of ``middle weight'', i.e., $35\le\Mtot \le 37.5 \Msun$ BBHs,
for a three-way comparison of \TEO{} with \NR, \Phn, and \SEOB.
We then used the additional $ 1030$ more inspirals
for an extended comparison with \SEOB{},
which comprised of $100 $ cases with BNS-like masses,  
a dozen cases with masses appropriate for black hole neutron star systems, 
approximately another 100 cases where one or both masses are in the lower mass gap,
i.e., $\lesssim 5 M_\odot$ \cite{Ozel:2010su, Farr:2010tu, Abbott:2020gyp},
and the remaining cases involving typical stellar-mass BBHs.
The non-precessing-binary parameters $\{f_0,\Mtot,q\}$ 
corresponding to these 1230 cases are shown in Fig.~\ref{fig:parameter_space},
where we additionally show $\{\chi_{\text{eff},0},\Sperp,\chi_p\}$ which
project the remaining five spin degrees of freedom, $\Sa(0),\Sb(0)$,
to just two (recall we set $\mbf{S}_{1y}(0)=0$).

To assess \TEO, we first computed frequency-domain matches
between \TEO-generated detector strains and those generated by $\{$\NR, \SEOB, \Phn$\}$ 
for the \nNR{} inspirals, then extended the match computation
to the expanded \TEO-\SEOB{} comparison set.
The match (or faithfulness) between two waveforms is computed by maximizing 
the following expression over initial time and phase shifts, $t_0, \phi_0$
\!\footnote{One can also maximize over $t_c,\phi_c$: time and phase shift at coalescence.}
\be
\M \equiv \max\limits_{t_0,\phi_0} \f{\langle h^\mathtt{k} | h^\mathtt{T}\rangle}{\sqrt{\langle h^\mathtt{k} | h^\mathtt{k}\rangle \langle h^\mathtt{T} | h^\mathtt{T}\rangle}}, \label{eq:Match}
\ee
where 
\be
\hspace{20mm} \langle h^\mathtt{k} | h^\mathtt{T}\rangle\equiv 4 \Re \int_{f_{k}}^{f_f}\f{\tilde{h}^\mathtt{k}(\fgw)\,\tilde{h}^{\mathtt{T}\ast}(\fgw)}{S_n(\fgw)} d\fgw
\label{eq:inner_prod}
\ee
is the inner product between the Fourier transforms of the GW strain,
$\tilde{h}^\mathtt{k},\tilde{h}^\mathtt{T}$, weighted by the one-sided power spectral density (PSD) $S_n(f) $ 
of the detector noise with \texttt{T} denoting \TEO{} and \texttt{k} = \NR, \Phn, \SEOB.
For the PSD, we use Advanced LIGO's ``zero-detuned high-power'' design sensitivity of Ref.~\cite{aLIGODesign_PSD}.
We set $f_i=1.05 f_0$, where recall $f_0$ is the initial non-precessing $(2,2)$ mode frequency.
As for $f_f$, since the current version of \TEO{} does not include ringdown, we opted for a suitable cutoff that is near the
peak of the twisted (2,2) mode, but slightly less:
$f_f=0.95f^{22}_\text{peak}$, to err on the side of caution.
There are many subtleties and complications in selecting
the proper peak when non-precessing modes first get ``mixed up'' in
the twist formula, after which the resulting precessing modes further
get mixed up in the mode sum \eqref{eq:hpc} for the GW strain.
As Ref.~\cite{Babak:2016tgq} discusses in their App.~D,
there may be cases in which several local peaks may be found, 
or none at all. 

In the time domain, the GW strain in a detector reads
\begin{align}
h(t) =  &F_{+} (\theta_s,\phi_s,\psi_s) h_{+}(t,\iota_s,\varphi_0)\nn\\ & + F_{\times} (\theta_s,\phi_s,\psi_s) h_{\times}(t,\iota_s,\varphi_0)\label{eq:strain} ,
\end{align}
where $F_{+,\times}$ are the detector
antenna pattern functions given by
\begin{align}
F_{+} (\theta,\phi,\psi)  &= \f{1+\cos^2\theta}{2} \cos2\phi \cos 2\psi -\cos\theta\sin2\phi\, \sin 2\psi \label{eq:Fplus}, \\
F_{\times} (\theta,\phi,\psi)  &=  \f{1+\cos^2\theta}{2}  \cos2\phi \sin 2\psi +\cos\theta\sin2\phi\, \cos 2\psi \label{eq:Fcross}.
\end{align}
In Eq.~\eqref{eq:strain}, $\theta_s,\phi_s$ are the sky-position angles, and $\psi_s$ is the polarization angle
of the GWs in the detector frame.
$h_{+,\times}$ are the standard GW polarizations which come from the following mode sum
\be
h_+ -i h_\times= \f{1}{D_L}\sum_{\el=2}^\infty\sum_{m=-\el}^\el h_{\el m}(t)\, {}_{-2}Y^{\ell m}({\iota_s},\varphi_0) \label{eq:hpc},
\ee
where $D_L$ is the distance to the source, which we set to $ 100\,$Mpc, 
${}_{-2}Y^{\ell m} $ are the (spin = $-2$)-weighted spherical 
harmonics, 
$\iota_s $ is the orbital inclination,
and $\varphi_0$ is the azimuthal angle
between the $x$-axis of the $\LN(0)$
frame and the projection of
the detector line-of-sight vector onto the plane
perpendicular to $\LN(0)$ ({see Fig.~\ref{fig:Frames}}).
Formally, $h(t)$ is obtained from a sum over all $\ell,m$ modes, 
but here, we suffice with the $\ell=2$ mode.
We will incorporate the available $\ell >2$ modes, which recently got upgraded \cite{Nagar:2020pcj}, in the next version
of \TEOB. Note that EOBNR approximants do not model $m=0$ modes,
so we set the (2,0) mode equal to zero.

%
%
%

\begin{widetext}

\begin{figure}  
      \includegraphics[width=0.99\textwidth]{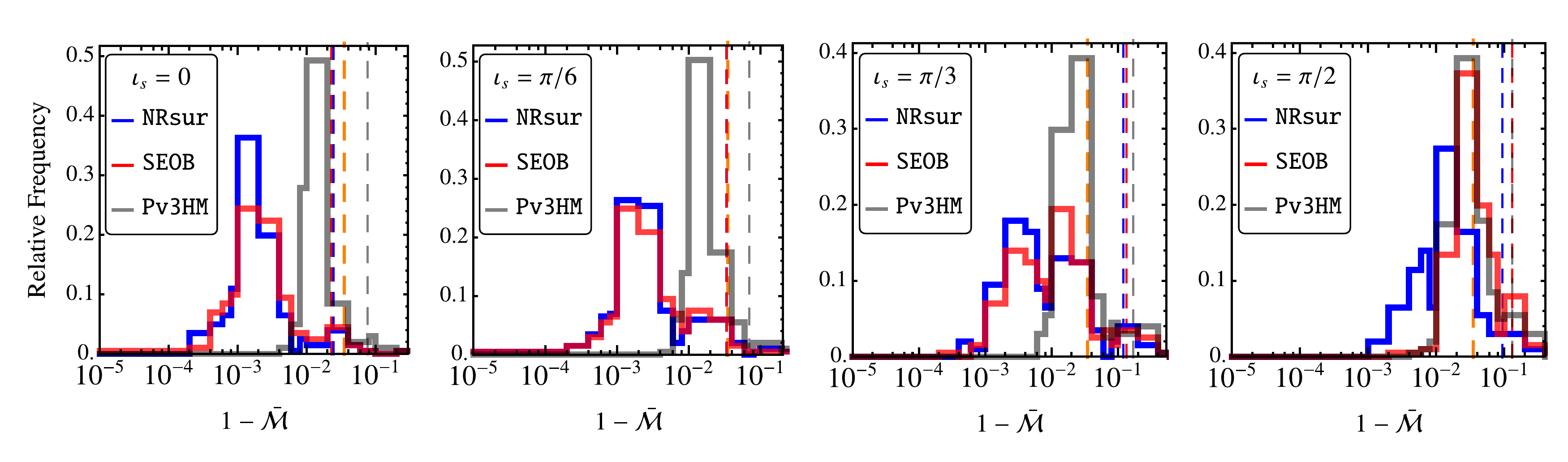}  
      \caption{\label{fig:Matches_200cases_3angle_av}
      The distribution the three-angle ($\psi_s,\theta_s,\phi_s $) averaged mismatch, $1-\Mbar$,
      between \TEOB{} and \NRsurP{} (blue), \SEOBNR{} (red), and
      \Phn{} (grey) for the 200 precessing inspirals of Sec.~\ref{sec:200case_summary} for orbital inclinations of $\iota_s=0,\pi/6,\pi/3,\pi/2$.
      The vertical orange dashed line marks the mismatch corresponding to
      $0.035$. The dashed blue, red, and black vertical lines mark the
      95th percentile of each set. 
      }
\end{figure}

\end{widetext}
It has become standard in waveform comparisons to use $\M=0.965$ as a benchmark.
This cutoff translates to the loss of roughly 10\% of events due to waveform systematics  \cite{Owen:1995tm, Flanagan:1997sx}.
We also employ this threshold 
and its mismatch counterpart $1-\M=0.035$
which we plot either as a horizontal or vertical dashed orange line in many of our subsequent figures.
%
\subsection{Summary of the main comparisons}\label{sec:200case_summary}
For our main comparison, we considered a set of 200 precessing
compact binary inspirals plotted as the blue-black dots
in the parameter space figure \ref{fig:parameter_space}.
For each inspiral we compared \TEO{} to
$\{$\NR,\,\SEOB,\,\Phn$\}$ by computing the detector strain
matches using Eq.~\eqref{eq:Match}.
As \NR{} has been shown to be better than 99\% faithful to NR
simulations for $> 95\%$ of the cases in its extrapolation space
\cite{Varma:2019csw}, it has become the current gold standard.
Therefore, we picked the parameters for our 200 cases to be well within
\NR's domain of interpolation, i.e., $1/4 \le q\le 1$ and
$\chi_1,\chi_2 \le 0.8$ \cite{Varma:2019csw}.
In order to maximize the number of orbital cycles, hence
the number of precession cycles, 
we set $\fgwi \in[35,40]\,$Hz and $\Mtot \in [35,40] \Msun$. 
Making these values any smaller tended to hit the low frequency bound of \NR, 
and setting them higher would miss the one, or at best two, precession cycles
that we expect. 
We further set $\chi_1=\chi_2=0.75$ since higher spins tend to 
lead to more pronounced precession, thus posing a tougher challenge
for the precessing approximants.

For the $h(t)$ computation in Eq.~\eqref{eq:strain} ,
we used a grid of $\iota_s=\{0,\pi/6,\pi/3,\pi/2 \}$ and
$\psi_s=\{0,\pi/8,\pi/4, 3\pi/8 \}$ with random values assigned for $\varphi_0 \in [0,2\pi)$ 
at each value of $\{\iota_s,\psi_s\}$.
For the sky angles $\{\theta_s,\phi_s\}$,
we employed a grid with spacing ${\pi/4}$. 
At each point in the four-angle grid,
we generated $h(t)$ using 
\TEO, \NR, \SEOB, and \Phn. 
This resulted in a total of $4\times 200 \times 4\times 4\times 4\times 8\sim 4\times10^5 $
strains from which
we computed the matches between \TEO{} and
$\{$\NR,\,\SEOB,\,\Phn$\}$ via Eq.~\eqref{eq:Match}
using the \texttt{Python} library \texttt{PyCBC} \cite{pycbc}.

For the $\sim 10^5$ \TEO-\NR{} matches,
we found that {91\%} were greater than 0.965 and less than 3\% of the sample
yielded $\M <0.9$ the majority of which happened with inclinations of $\iota=\pi/3$ and $\pi/2$.
Similarly, {85}\% of the \TEO-\SEOB{} matches 
and {77}\% of the \TEO-\Phn{} matches were greater
than 0.965. These percentages remained within $\pm 1\%$ 
when we switched from an evenly spaced
$\{\theta_s,\phi_s\}$ grid to a random one as well as when we repeated the entire computation
with new random values for $\varphi_0$.
To summarize our main results, we introduce the three-angle averaged match,
$\Mbar$ as follows. 
Given a set of binary parameters, we fix $\iota_s,\varphi_0$
then compute the match $\M_{ijk}$ between a given pair of approximants at each
of the $4\times4\times8=128$ points in the $\{\psi_{si},\theta_{sj},\phi_{sk}\}$
grid. $\Mbar$ is then just the straightforward discrete mean of $\M_{ijk}$.
Note that since by definition $0\le \M \le 1$ and
our main threshold is $\M=0.965$, the averaging tends to produce
lower percentages of $\Mbar >0.965$ cases. Therefore, we present
percentages over our entire match set, but use $\Mbar$ in our figures.

In Fig.~\ref{fig:Matches_200cases_3angle_av} we present
the distributions of the three-angle averaged mismatch, $1-\Mbar$, 
between \TEO{} and the validation approximants \NR, \SEOB, and \Phn{}
for $\iota_s = \{0, \pi/6,\pi/3,\pi/2\}$.
As can be seen in the figure, the majority of the
mismatches lays to the left of $0.035$ represented by the vertical dashed orange line.
The vertical dashed $\{$blue, red, gray$\}$ lines respectively represent the 95th percentile
\TEO-$\{$\NR,\,\SEOB,\,\Phn$\}$ mismatches.
Clearly, for $\iota \le \pi/6$, \TEO{} matches \NR{} and \SEOB{}
better than 0.965 for more than 95\% of the cases.
For the \TEO-\Phn{} matches, this is roughly 86\%. 
The shift of the peak of $1-\Mbar$ from $10^{-3}$ to $10^{-2}$
as $\iota_s$ increases is also evident in the figure for the \TEO-\NR{} (blue) and
\TEO-\SEOB{} (red) histograms,
whereas for the \TEO-\Phn{} distribution (gray) this shift is much less
pronounced with the peak of the distribution also remaining much narrower
\begin{widetext}

\begin{figure}[t]
\hspace{-6mm}  
   \includegraphics[width=1.02\textwidth]{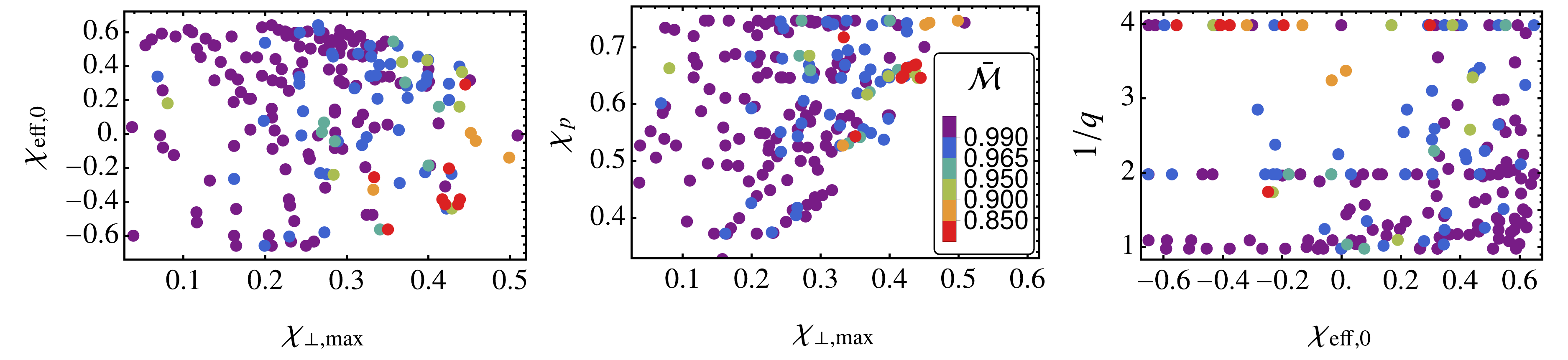}
\caption{\label{fig:NRsur_match_corner_plots} 
Two-dimensional scatter plots of the three-angle averaged match, $\Mbar$, between \TEOBResumP{} and \NRsurP{} for the \nNR{} precessing compact binary inspirals as ``seen'' at an orbital inclination of $\iota_s=\pi/3$. 
For this figure, we opted for $1/q$ to better relate to values
more familiar in the NR community.
The cooler colors (purple, blue) represent cases with $\Mbar \ge 0.965$ 
while the warmer colors (orange, red) represent 
$\Mbar \le 0.9$.
See Sec.~\ref{sec:twist_NRsur} for details as to why the match degrades in certain regions.
The distribution of the colors is roughly the same for $\iota_s=\pi/6,\pi/2$,
albeit with very few ``hot'' dots for the former and about the same number for
the latter. 
}
\end{figure}
\end{widetext}
for all inclinations.

The deterioration of \TEO's agreement with the other approximants 
for increasing $\iota_s$ is expected since 
the precessing $(2,\pm1), (2,0)$ modes contribute more to the 
GW strain as $\iota_s$ increases.
The disagreements in these modes stem from disagreements in the 
\emph{non-precessing} $(2,\pm1), (2,0)$ modes.
For example, we found that while all, but one, non-precessing $(2,2)$ modes of \texttt{TEOBResumS} matched their non-precessing 
\texttt{NRSur} counterparts to better than 0.99, only 60\% of the non-precessing $(2,1)$ modes achieved matches greater than $0.965$\footnote{We employed the \texttt{gwsurrogate}
package \cite{gwsurrogate, Field:2013cfa} to generate
the (non)precessing $\ell=2$ modes of \texttt{NRSur}.}.
We further confirmed that most of the worst strain mismatches do indeed come
from cases where the non-precessing 
$(2,1)$ mode matches between \TEO{} and
the validation approximants are less than 0.9.
This is consistent with the findings of Ref.~\cite{Ramos-Buades:2020noq},
where the effects of mismodelling the non-precessing $(2,1)$ and higher
modes were systematically investigated.
As expected, we also found out that the worst matching cases
between \TEO{} and the validation approximants have 
$0.65 \lesssim \chi_p \le 0.75$ and
$q\lesssim 1/3$ with the majority having $q=1/4$.

Overall, for the set of precessing compact binary inspirals considered 
in this section with $0.3\lesssim \chi_p \le 0.75$,
\SEOB{} showed the best agreement with \NR{}
with only 6.9\% of the $\sim 10^5$ matches below 0.965. 
This percentage was $14\%$ for \NR-\Phn{} and 18\% for \SEOB-\Phn{} matches.
For the inclination of $\iota_s=\pi/2$, \Phn{} matched
\NR{} best with 91\% of the cases yielding $\M>0.95$, 
whereas \TEO{} and \SEOB{} had 88.6\% and 86.6\% of these cases yield $\M>0.95$, respectively. These differences once again highlight the importance of having
several different waveform approximants.
We present additional details of \TEO's performance against $\{$\NR,\,\Phn,\,\SEOB$\}$ in the next
subsections.


\subsection{Comparisons with \NRsurP{} waveforms}\label{sec:twist_NRsur}
For the $\sim 10^5$ \TEO-\NR{} matches that we computed,
we found that  74.1, 91.1, 93.8\% yielded $\M>0.99, 0.965, 0.95$, respectively.
Within the four $\iota_s=0,\pi/6,\pi/3,\pi/2$ subsets,
97.6, 95.5, 86.7, 84.7\% yielded $\M > 0.965$.
Similarly, 97.0\% of the $q<1/4$ matches gave $\M >0.965$
in contrast to 78.1\% of the $q=1/4$ cases.
Of the $q=1/4$ cases, about 94\% and 86\% of the $\iota_s=0,\pi/6$
subsets yielded $\M> 0.965$ as opposed to
only about 2/3 of the $\iota_s=\pi/3,\pi/2$ subsets
giving $\M > 0.965$.

The increase of mismatch with decreasing mass ratio and increasing orbital
inclination is a direct outcome of the increasing mismatch
between the precessing $(2,\pm1), (2,0)$ modes of \TEO{} and \NR.
This disagreement, in turn, stems mostly from the less-than-ideal
agreement between the non-precessing $(2,\pm1)$ modes \TEO{} and \NR{}
mentioned in Sec.~\ref{sec:200case_summary}.
Additionally, \TEO{} sets $h_{20}^\text{NP}=0$ which \NR{} does not,
but the mismatch due to this assignment is subdominant as
the amplitude of $h_{20}^\text{NP}$ is
orders of magnitude smaller than the amplitude of $h_{21}^\text{NP}$.
As a test, we replaced $h_{21}^{\text{NP},\mathtt{TEOB}}$ with $h_{21}^{\text{NP},\mathtt{NRsur}}$
for a few cases and observed that the resulting \TEO-\NR{} match improved, verifying our above hypothesis that the non-precessing $(2\pm1)$
modes are mostly responsible for the high-inclination, low-$q$ mismatches.

We also explored how the \TEO-\NR{} match 
behaves across the precessing binary parameter space.
In Fig.~\ref{fig:NRsur_match_corner_plots},
we show two-dimensional scatter plots of the match 
against $\chi_{\text{eff},0}, \chi_p,\Sperp$, and $1/q$ for $\iota_s=\pi/3$.
As can be deduced from the middle panel of the figure,
the match worsens for larger values of $\chi_p, \Sperp$, i.e.,
stronger precession, indicated by 
the ``warmer'' colors (red, orange).
Also evident in the right panel is the aforementioned
degradation of the match for the $q=1/4$ cases.
Interestingly, the match also worsens
for more negative values $\chi_{\text{eff},0}$
hinted both in the left and right panels.
This again relates back to the mismatch in the non-precessing
$(2,\pm1)$ modes.
These trends persist for $\iota_s=\pi/6, \pi/2$, 
albeit less pronounced for the former.

%
\begin{figure}[t]
\centering
      \includegraphics[width=0.49\textwidth]{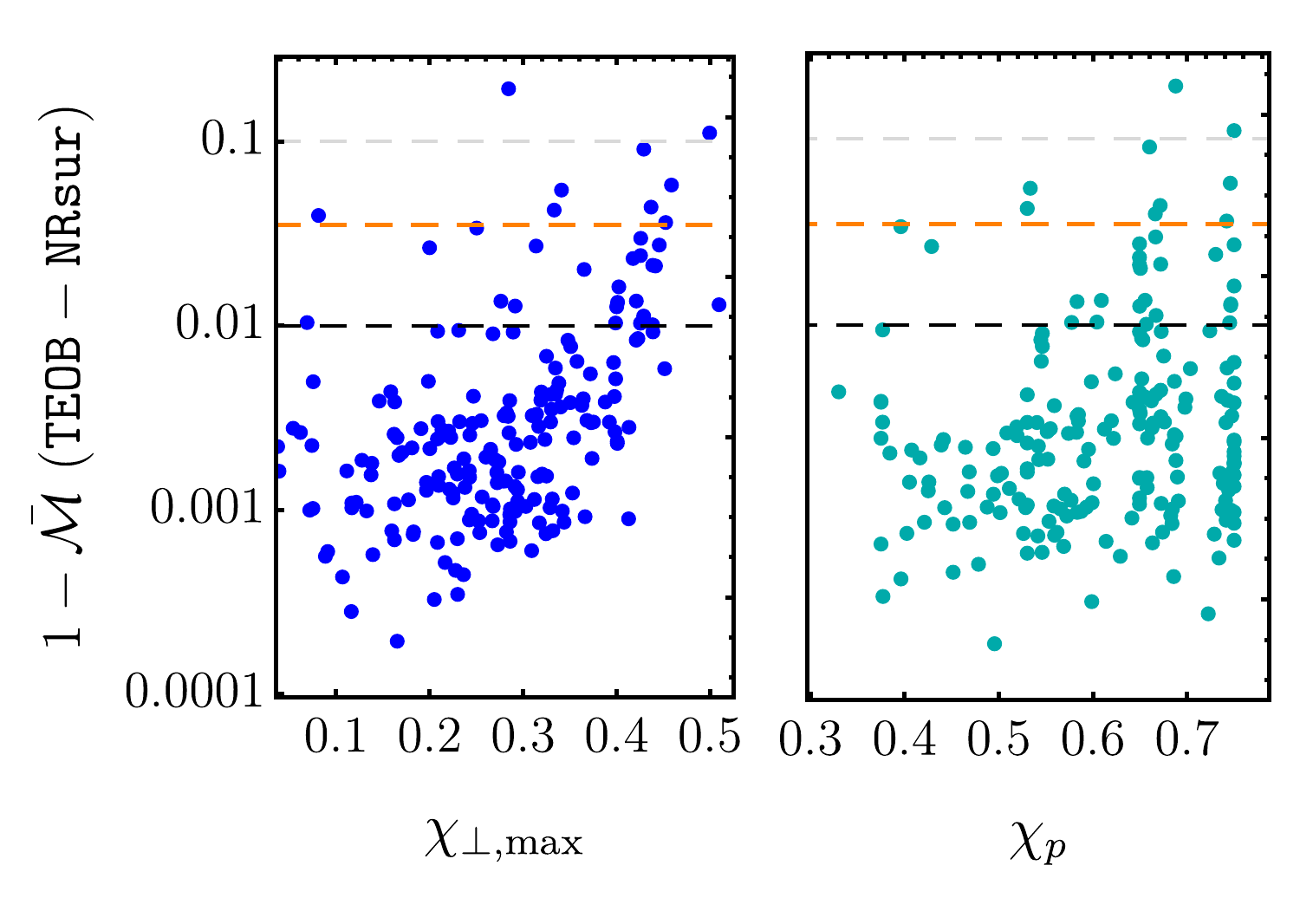}  
\caption{\label{fig:NRsur_TEOB_mismatches_vs_SperpMax} 
Three-angle averaged mismatch, $1-\Mbar$ between \TEOBResumP{} and \NRsurP{} vs $\Sperp$ (left panel, blue dots) and $\chi_p$ (right panel, teal dots)
for $\iota_s=\pi/6$.
The horizontal dashed black, orange, and gray lines represent 
mismatches of 0.01, 0.035, and 0.1, respectively.
The mismatch seems to depend more strongly on $\Sperp$ 
than $\chi_p$, which suggests that $\Sperp$ may somehow
expose a systematic error in \TEOB{} resulting from
twisting constant-spin non-precessing waveforms as opposed to
time-varying ones.
}
\end{figure}

As an interesting side note, we compared in Fig.~\ref{fig:NRsur_TEOB_mismatches_vs_SperpMax}
how $1-\Mbar$ changes when plotted against $\chi_p$ versus against $\Sperp$
for $\iota_s=\pi/6$.
The [semilog] plots hint that $1-\Mbar$ shows a vague exponential dependence on $\Sperp$,
but not on $\chi_p$.
This trend persists for other inclinations, albeit with more
outliers for larger values of $\iota_s$. 
The trend also shows in the mismatches of \TEO{} with \Phn{}
as we illustrate in the next subsection.
The trend even persists for \NR-\Phn{} and \NR-\SEOB{} mismatches,
albeit less clearly, but again more strongly for small $\iota_s$
as in \TEO-\NR{} mismatches.
This suggests that $\Sperp$ might somehow expose
a systematic error in the way that the approximants
generate their precessing waveforms.
A more detailed study is required to firmly
establish this (or refute it).
Nonetheless, based on these findings and given that the values for $\Sperp$ seem less degenerate than $\chi_p$ (at least for the 200 cases here),
we believe that $\Sperp$ may be useful in future parameter estimation
studies. In the least, it seems to encode the strength of precession.


\subsection{Comparisons with \PHM{} waveforms}\label{sec:twist_Phenom}

Of the $\sim 10^5$ \TEO-\Phn{} matches,
77.6\% are greater than $0.965$ and 86.4\% greater than $ 0.95$.
The percentage of matches greater than 0.965
in the four $\iota_s=0,\pi/6,\pi/3,\pi/2$ subsets are
91.4, 87.2, 71.1, 60.7, respectively.
As with the \NR{} comparisons, the $q=1/4$ subset has
fewer $\M> 0.965 $ cases, 45\% of the set,
than the $q<1/4$ subset, 87.6\%.
Within the $q=1/4$ subset, 65\% and 57\% of the $\iota_s=0,\pi/6$
subsets have $\M> 0.965$ as opposed to
only 36\%, 22\% for the $\iota_s=\pi/3,\pi/2$ subsets
(these last two percentages are greater than 50\%
when considering $\M=0.95$).

The way \TEO{} compares with \Phn{} is roughly consistent with the
way it compares with \NR, albeit with lower percentages of $\M>0.965$
cases overall and within the chosen subsets.
This consistency is evident when comparing
Fig.~\ref{fig:Pv3HM_match_corner_plots} with
Fig.~\ref{fig:NRsur_match_corner_plots}, i.e, the two-dimensional
scatter plots of $\Mbar$ for $\iota_s=\pi/3$.
In both figures, many of the red dots are located at the same positions in 
the $\{\chi_{\text{eff},0}, \Sperp, \chi_p, 1/q\}$ space,
with some orange dots of Fig.~\ref{fig:NRsur_match_corner_plots}
also having become red.
In fact, the major difference between the two figures is the
``reddening'' of the dots, consistent with 
Fig.~\ref{fig:Matches_200cases_3angle_av} where
the position of the peak of the distribution of \TEO-\NR{} mismatches
is roughly an order of magnitude smaller than the peak
of the distribution of \TEO-\Phn{} mismatches,
hence the domination of Fig.~\ref{fig:NRsur_match_corner_plots}
by the purple dots, and of Fig.~\ref{fig:Pv3HM_match_corner_plots}
by the blue dots.
We should re-emphasize that it is not just \TEO{} that produces
increasing mismatches for small $q$ and large $\chi_p, \iota_s$.
In fact, \Phn{} exhibits a similar degradation in its matches with \NR,
as does \SEOB{} (but less so). When compared with each other, all approximants show increasing disagreements
in this challenging region requiring excellent match of all precessing modes, not just the $(2,2)$ mode.

In Fig.~\ref{fig:Pv3HM_TEOB_mismatches_vs_SperpMax}
we plot the three-angle averaged mismatch between \TEO-\Phn{}
against $\Sperp$ and $\chi_p$ for $\iota_s=\pi/6$.
As in Fig.~\ref{fig:NRsur_TEOB_mismatches_vs_SperpMax},
a vague exponential relation between $1-\Mbar$ and $\Sperp$
can be discerned.
Analogous to the \TEO-\NR{} comparisons,
this relation persists for other values of $\iota_s$.
As we already discussed the implications of this relation
in the previous section, we move on to comparisons of \TEO{}
with \SEOB.


\subsection{Extensive comparisons with \SEOBNR{} waveforms}\label{sec:twist_SEOBNR}
\SEOBNR{} is the latest precessing approximant
within the \texttt{SEOBNR} family.
As the upgrade to \texttt{SEOBNRv3} \cite{Pan:2013rra, Babak:2016tgq},
it incorporates precession in higher modes up to $\el=5$ \cite{Ossokine:2020kjp}.
The precession in \SEOBNR{} (also in \texttt{v3})
is coupled to the aligned-spin EOB dynamics
so that the resulting aligned-spin waveforms in the co-precessing frame are obtained from time-dependent $\chi_1,\chi_2$. 
Most recent comparisons using
approximately 1500 precessing
SXS simulations have yielded
\SEOBNR-NR matches of $> 0.97$ for $> 94\%$ of the cases
with the higher modes included
\cite{Ossokine:2020kjp}. Note that these currently consist of only the
$(3,\pm3),(4,\pm4),(5,\pm5)$ modes lacking the important
and more challenging $(3, \pm2)$ and $(4, \pm3)$ modes \cite{Ossokine:2020kjp}.
Moreover, \SEOBNR{} is not yet calibrated to NR waveforms 
in the precessing sector, but only to the aligned-spin waveforms.
Nonetheless, along with \texttt{IMRPhenomXPHM}
\begin{widetext}

\begin{figure}[t]
\hspace{-6mm}  
   \includegraphics[width=1.02\textwidth]{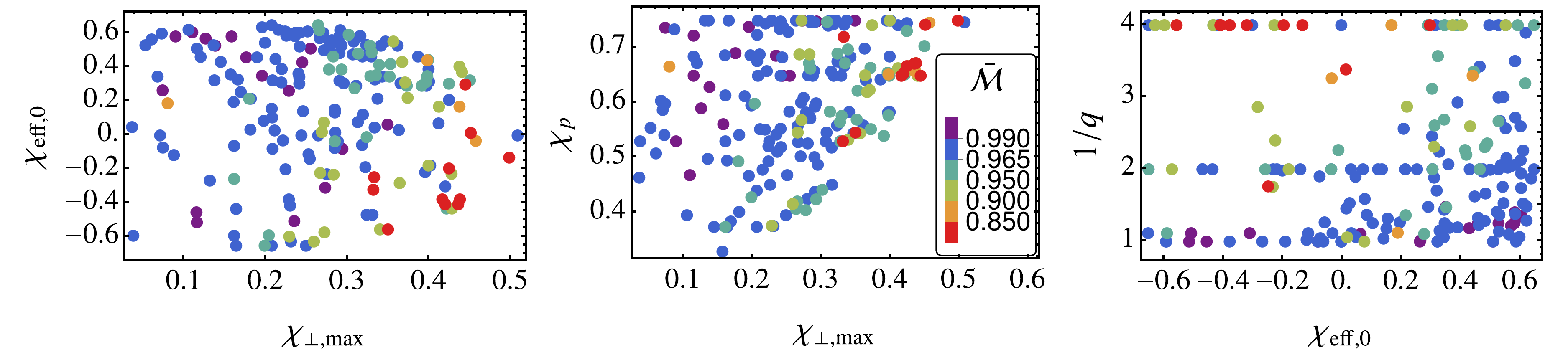}
\caption{\label{fig:Pv3HM_match_corner_plots} 
Same as Fig.~\ref{fig:NRsur_match_corner_plots}, but now for matches between \TEOBResumP{} and \PHM{} once again for $\iota_s=\pi/3$.  
Note that many of the red dots here are the same as those of Fig.~\ref{fig:NRsur_match_corner_plots}. This similarity persists for all 
values of $\iota_s$.
}
\end{figure}

\end{widetext}
\cite{Pratten:2020ceb}, \SEOBNR{} is currently one of the most NR-faithful,  
non-surrogate precessing approximants.

Since \SEOBNR{} does not suffer from the current parameter limitations
of \NRsurP{}, we used a larger set of 1230 precessing inspirals with the parameters spanning greater ranges.
In particular, for the key parameters, we have:
$ 0\le \chi_p\le 0.993, f_0 \ge 20\,\text{Hz}, 
0.1 \le q \le 1$, and $ 3 \Msun \le \Mtot \le 70 \Msun$ (see Fig.~\ref{fig:parameter_space}). We realize that comparing cases with
$\chi_p$ in excess of 0.9 is rather ambitious, especially since
\SEOB{} has been tested against NR only up to this limit \cite{Ossokine:2020kjp}. Nonetheless, Ref.~\cite{Ossokine:2020kjp} also 
presented an \SEOB-\Phn{} comparison up to $\chi_p \lesssim 0.99$
so we proceed in the same spirit.

Within this expanded set, 200 cases have already been partly 
discussed in Secs.~\ref{sec:200case_summary},
where we reported the \TEO-\SEOB{} matches and their distribution
in Fig.~\ref{fig:Matches_200cases_3angle_av}.
Here, we add to this an expanded set of 1030 cases
for which we once again computed the $\{\psi_s,\theta_s,\phi_s\}$-averaged
matches, $\Mbar$, between \TEO{} and \SEOB{} for inclinations of $\iota_s=0,\pi/6,\pi/3$. As we discuss below, we leave the $\iota_s=\pi/2$ comparison
to future work.
As before, we used a $4\times4\times8$ grid for $\{\psi_s,\theta_s,\phi_s\}$
while assigning random values to $\varphi_0$.
This amounted to $128\times 1030\approx 1.3\times 10^5$ matches
computed for each inclination.
We set $D_L=100\,$Mpc as before.

%
\begin{figure}
\centering   
      \includegraphics[width=0.49\textwidth]{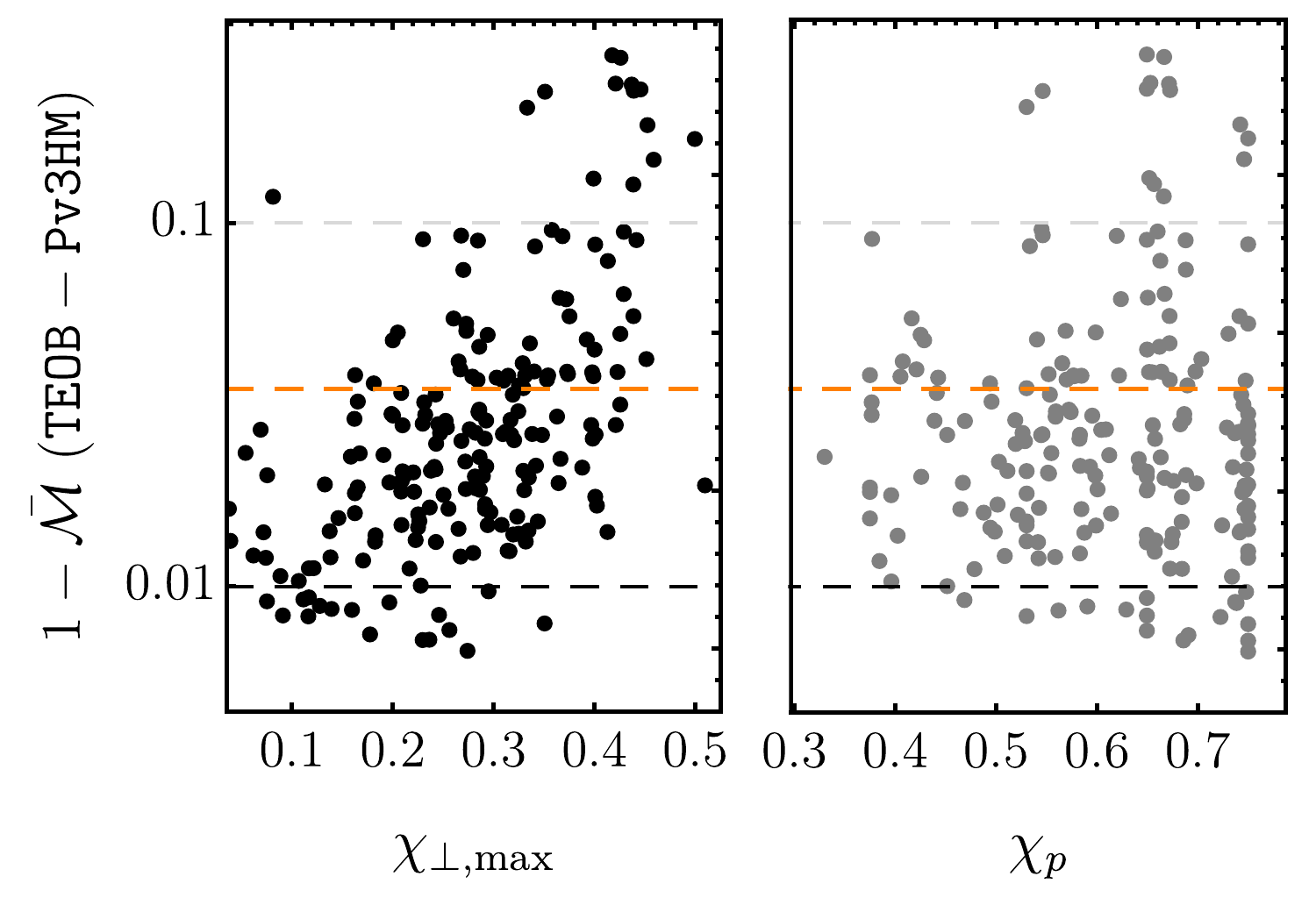}  
\caption{\label{fig:Pv3HM_TEOB_mismatches_vs_SperpMax} 
The three-angle averaged mismatch, $1-\Mbar$ between \TEOBResumP{} and \PHM{} plotted against $\Sperp$ and $\chi_p$ for $\iota_s=\pi/6$.
As in Fig.~\ref{fig:NRsur_TEOB_mismatches_vs_SperpMax}, 
the mismatch increases with increasing $\Sperp$ (black dots, left panel),
but seems to depend less strongly on $\chi_p$ (gray dots, right panel).
The horizontal dashed black, orange, and gray lines represent $1-\Mbar=0.01, 0.035, 0.1$, respectively.
}
\end{figure}

For the full set of 1230 cases, 90\% of the $\iota_s=0,\pi/6$ matches are above 0.965
with this percentage dropping to 75\% for $\iota_s=\pi/3$.
We checked that these percentages remained unchanged (to less than 0.5\%) 
when using randomly assigned values for $\theta_s, \phi_s$ instead of a grid
with spacing of $\pi/4$.
Part of the reason for the increased disagreement with respect to the 
\TEO-\NR{} comparison is the fact that now roughly 7.5\% of the 1230
\emph{non}-precessing \TEOBResumS-\SEOB{} (2,2)-mode matches are less than 0.965,
whereas there was a single non-precessing \TEOBResumS-\NR{} (2,2) mode
match less than 0.99 out of 200 cases.
Some of this (2,2)-mode disagreement is due to the increased range of $q$ down to 0.1, for which
we find that there are indeed increased occurrences of non-precessing
(2,2) mode matches less than 0.965 for $q\lesssim 0.2$.
Moreover, 42\% of the non-precessing (2,1) mode matches are also less than
0.965. This latter disagreement manifests a more prominent mismatch in the
precessing $(2,\pm1), (2,0)$ modes which matter more
for cases with strong precession and larger inclination.
Therefore, given that nearly 60, 25\% of the 1230 cases have
$\chi_p \ge 0.5, 0.7$ with a mean of 0.55, the degradation we observe in 
$\Mbar$ when going from $\iota_s=0,\pi/6$ to $\iota_s=\pi/3$ is not surprising. 
A similar disagreement has been shown between \SEOB{} and \Phn{} 
for $\chi_p\gtrsim 0.7$ at $\iota_s=\pi/3$ \cite{Ossokine:2020kjp}, 
but a similar $\iota_s=\pi/2$ comparison was not reported there.
For our set, we find that $\Mbar$ degrades even more severely
when going from $\iota_s=\pi/3$ to $\pi/2$ with only half the matches
greater than 0.85. Again the culprit mostly seems to be the precessing
(2,0) mode for which the \TEO-\SEOB{} matches are mostly in the range of
0.6 to 0.8. As this requires further investigation, we limit our comparisons
here to $\iota_s \le \pi/3$.

\begin{widetext}

\begin{figure}  
      \includegraphics[width=0.99\textwidth]{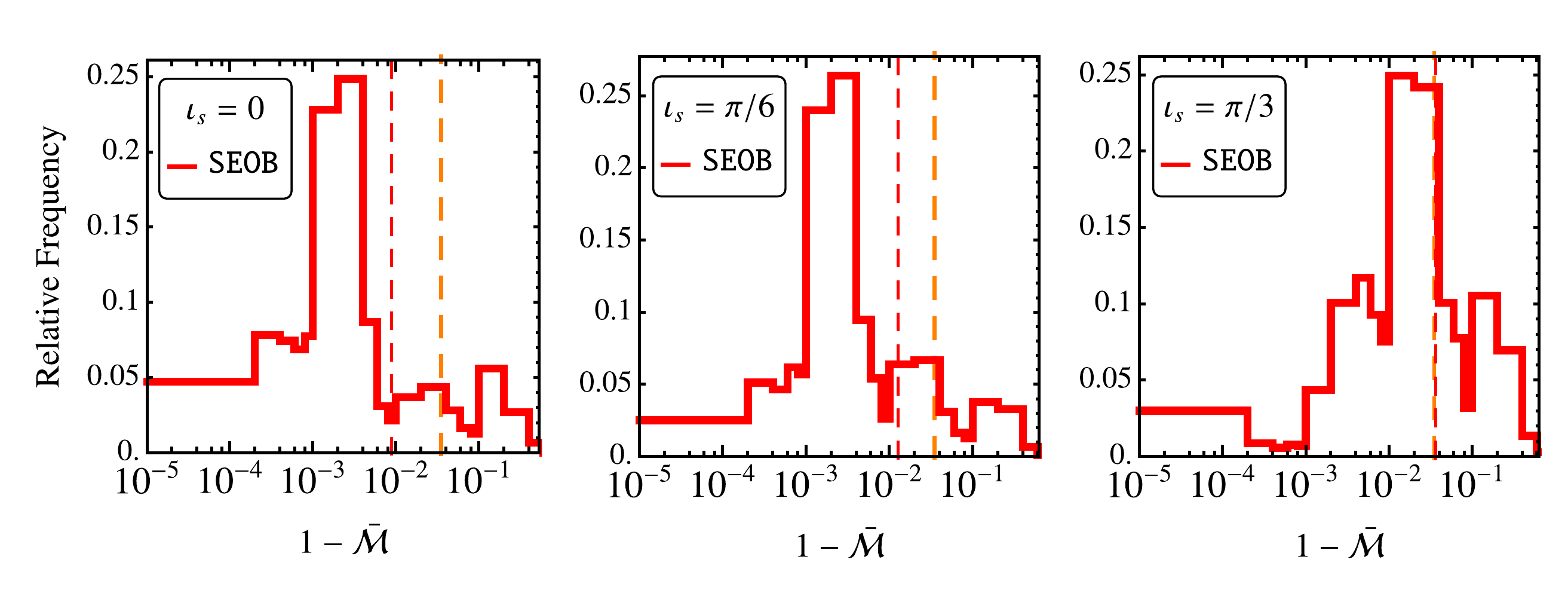}  
      \caption{\label{fig:Matches_1036set_3angle_av}
      The distribution the three-angle ($\psi_s,\theta_s,\phi_s $) averaged mismatch, $1-\Mbar$,
      between \TEOB{} and \SEOBNR{} for the additional 1030 precessing inspirals for orbital inclinations of $\iota_s=0,\pi/6,\pi/3$.
      The vertical orange dashed line marks the mismatch corresponding to
      $0.035$. The vertical dashed red line marks the
      95th percentile. The parameters for the 1030 cases are 
      represented by the red dots in Fig.~\ref{fig:parameter_space} 
      that do not overlap with the blue, black markers.}
\end{figure}

\end{widetext}

In Fig.~\ref{fig:Matches_1036set_3angle_av}, we show the distribution
of the \TEO-\SEOB{} three-angle-averaged mismatches, $1-\Mbar$, 
%
%
%
%
%

%
%
%

\begin{figure}[t]
   \includegraphics[trim=0 0 0 0,clip,width=0.5\textwidth]
   {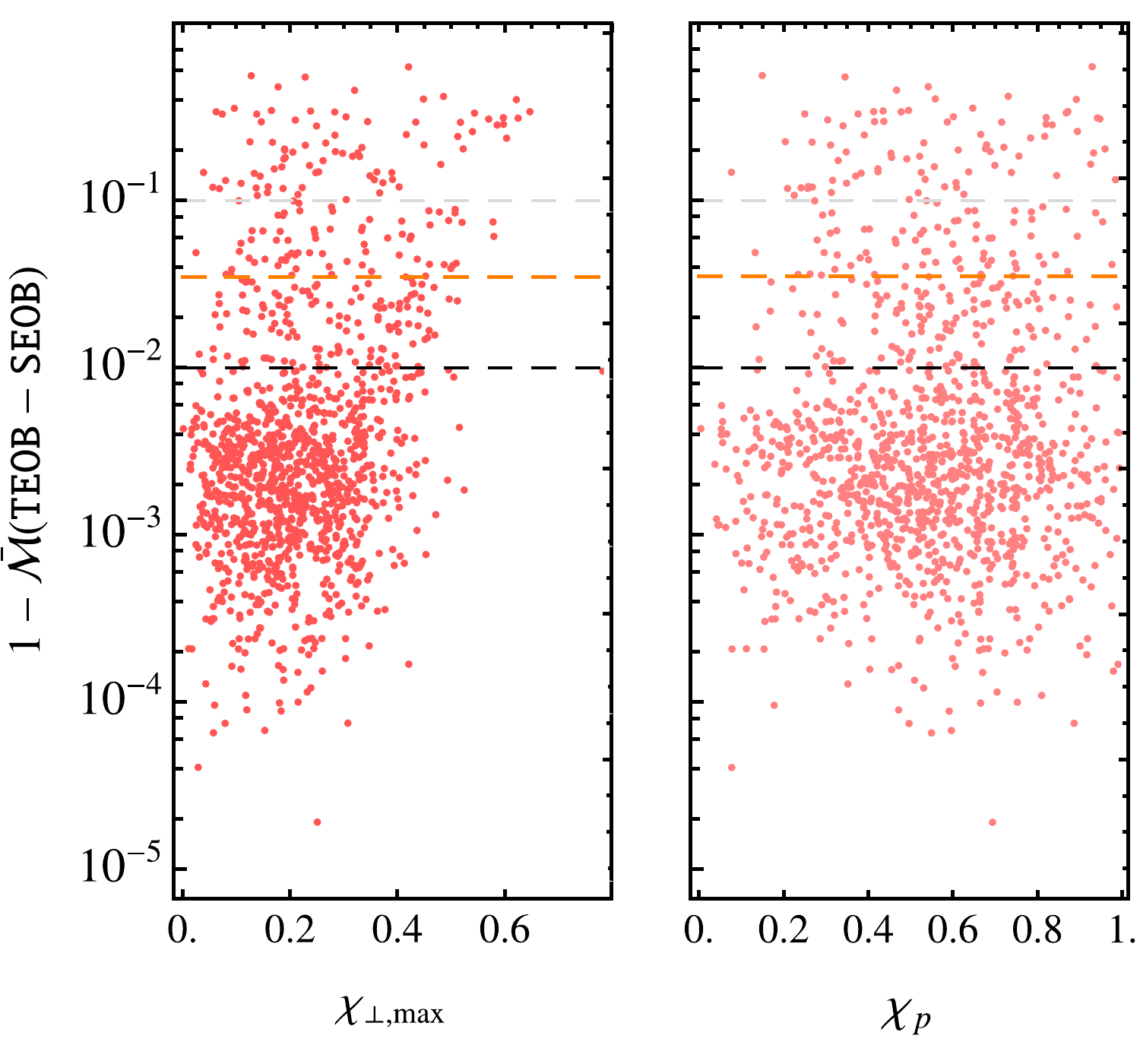}
\caption{\label{fig:SEOB_Match_vs_Sperp} 
The three-angle averaged mismatch, $1-\Mbar$ between \TEOBResumP{} and \SEOBNR{} plotted against $\Sperp$ (red) and 
$\chi_p$ (pink) for $\iota_s=\pi/6$.
Though the trend seen in Figs.~\ref{fig:NRsur_TEOB_mismatches_vs_SperpMax} and \ref{fig:Pv3HM_TEOB_mismatches_vs_SperpMax}
with respect to $\Sperp$ seems to have gotten mostly
``buried'', its plot still looks like less of a random scatter than 
the corresponding $\chi_p$ plot.
The horizontal dashed black, orange, and gray lines represent $\Mbar=0.99, 0.965, 0.9$, respectively.}
\end{figure}
\noindent
for the three inclinations.
As can be seen in the figure, for $\iota_s=0$ and $\pi/6$,
the mismatches have a tall, narrow distribution centered at roughly
$2\times 10^{-3}$, which becomes broader and shifts to roughly
$2\times 10^{-2}$ for $\iota_s=\pi/3$.

We also checked whether or not the $\ln(1-\Mbar)$ vs $\Sperp$ trend 
of Figs.~\ref{fig:NRsur_TEOB_mismatches_vs_SperpMax} and \ref{fig:Pv3HM_TEOB_mismatches_vs_SperpMax} 
persisted for the entire set of 1230 \TEO-\SEOB{} matches, 
which we show in Fig.~\ref{fig:SEOB_Match_vs_Sperp} for $\iota_s=\pi/6$, 
where we also plot $\ln(1-\Mbar)$ vs $\chi_p$ as before.
It is clear from the new figure that the vague trend we had previously
discerned has more or less disappeared as the set size increased
by roughly an order of magnitude as well as the range of $q, f_0, \Mtot$.
This is not unexpected since more cases with greater range of parameters
may increase the potential causes of disagreement between waveform approximants, thus burying the rough trend of  
Figs.~\ref{fig:NRsur_TEOB_mismatches_vs_SperpMax} and \ref{fig:Pv3HM_TEOB_mismatches_vs_SperpMax}.
Indeed, the alternate version of Fig.~\ref{fig:SEOB_Match_vs_Sperp}
made using only the 200 cases of Secs.~\ref{sec:200case_summary}-\ref{sec:twist_Phenom} looks very similar to 
Figs.~\ref{fig:NRsur_TEOB_mismatches_vs_SperpMax} and \ref{fig:Pv3HM_TEOB_mismatches_vs_SperpMax}.
The proper way to check for this trend 
is to compare precessing waveforms for only the cases for which the non-precessing
modes show excellent agreement (e.g., matches $>$ 0.99) 
then slowly increase $\theta_1,\theta_2$
while keeping all other parameters unchanged, thus only
increasing $\Sperp$ and $\chi_p$. 
The resulting plots of $\ln(1-\Mbar)$ vs $\Sperp$ and $\chi_p$
would be much more conclusive as to whether or not the trend with
respect to $\Sperp$ exists. 
We leave this for future work.

\begin{widetext}

\begin{figure}[t]
\centering   
   \includegraphics[width=0.99\textwidth]{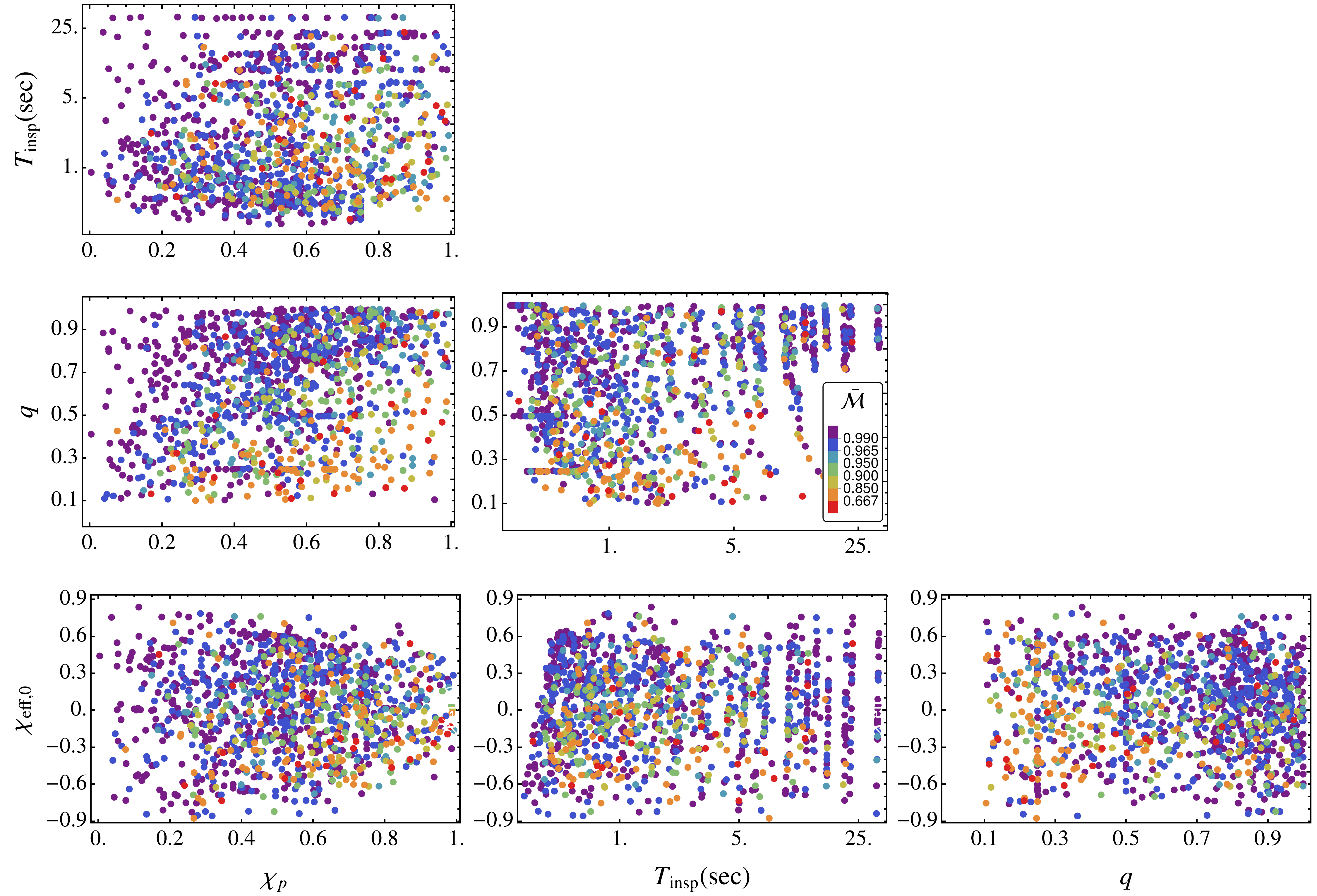}
\caption{\label{fig:SEOB_Match_Corners} 
Similar to Figs.~\ref{fig:NRsur_match_corner_plots} and \ref{fig:Pv3HM_match_corner_plots},
but now for matches at an inclination of $\iota_s=\pi/3$
between \TEOBResumP{} and \SEOBNR{} for the entire set of 1230 cases.
We have also added the inspiral time, $T_\text{insp}$ (note the log scale) to the plots.
As written in Sec.~\ref{sec:twist_SEOBNR}, the matches degrade for
low values of $q$, more negative values of $\chi_{\text{eff},0}$, 
and increasing values of $\chi_p$ (or $\Sperp$).
The corresponding figures for $\iota_s=0,\pi/6$ are similar, albeit
with fewer red and orange dots, and more purple dots.
}
\end{figure}
\end{widetext}

Since we greatly expanded the ranges of a few parameters,
we investigated how this may affect the \TEO-\SEOB{} matches
by plotting them against $\chi_p,q,\chi_{\text{eff},0}$, and
the inspiral time $T_\text{insp}$,
in Fig.~\ref{fig:SEOB_Match_Corners} for $\iota_s=\pi/3$.
As in Figs.~\ref{fig:NRsur_match_corner_plots}, \ref{fig:Pv3HM_match_corner_plots} 
we observe increasing mismatches for larger values of 
$\chi_p$ (and $\Sperp$) and more negative values of $\chi_{\text{eff},0}$.
Additionally, the matches worsen for $ q\lesssim 0.25$.
This is not unexpected as it is known that 
the mismatch between the non-precessing \TEOBResumS{} and \texttt{SEOBNRv3}
increases as $q$ decreases \cite{Nagar:2018zoe},
but a similar investigation between \TEOBResumS{} and \texttt{SEOBNRv4}
has not yet been conducted. However, we have already mentioned
that we have observed some $q\lesssim 0.2$ cases with the \TEOBResumS-\SEOB{}
non-precessing (2,2) mode matches less than 0.965.

The inspiral time seems to have no affect on the matches,
up to the longest inspirals considered here, i.e., 25\,seconds.
The corresponding plots for $\iota_s=0,\pi/6$ contain the same regions
of degrading matches, albeit with very few orange and red dots.
Combining Figs.~\ref{fig:NRsur_match_corner_plots}, \ref{fig:Pv3HM_match_corner_plots} and \ref{fig:SEOB_Match_Corners}, 
we can conclude that the most
challenging ``corner'' of the parameter space for \TEOB{} to match other precessing approximants is the three-dimensional $q \lesssim 0.25, \chi_{\text{eff},0}\lesssim -0.5, \chi_p \gtrsim 0.6$ region.
The small-$q$, large-$\chi_p$ corner also seems to be a region of
increased mismatch between \SEOB{} and \Phn{} as shown in Fig.~14 of 
Ref.~\cite{Ossokine:2020kjp} and also between \Phn{} and NR as
hinted by Ref.~\cite{Khan:2019kot} though there
were only three NR simulations for the comparison.
Increasing mismatches for larger $\chi_p$ values and $q\le 1/5$
have also been observed between \texttt{IMRPhenomXPHM} and NR
simulations \cite{Pratten:2020ceb}.

\section{Conclusions} \label{Sec:end}
In this article, we introduced \TEOB: the precessing upgrade to \TEOBResumS{}.
Currently, \TEOB{} generates precessing $\ell=2, m\in[-2,2]$ modes by Euler-rotating
non-precessing (aligned, constant spin) \TEOBResumS{} modes from 
the instantaneous, non-inertial $\LN(t)$ frame to the inertial $\LN(0)$ frame.
This frame rotation, given by Eq.~\eqref{eq:hlm_Twist2},  is performed with Wigner's D matrices.
As it is, \TEOB{} generates precessing modes only up to merger
taken to be the peak of the twisted (2,2) mode.

We assessed the faithfulness of \TEOB{} by computing 
the polarization-declination-right-ascension averaged 
$\ell$=2 detector strain matches between \TEOB{} and $\{$\NRsurP{}, \texttt{IMRPhenomPv3HM}, \SEOBNR$\}$
for \nNR{} binaries at orbital
inclinations of $\iota_s=0, \pi/6,\pi/3$, and $\pi/2$.
We further compared \TEOB{} against \SEOBNR{} for
an additional set of 1030 binaries.

We also introduced a new parameter, 
$\Sperp$, in Eq.~\eqref{eq:S_perpMax},
which encodes the strength of precession.
We showed in Secs.~\ref{sec:twist_NRsur}-\ref{sec:twist_SEOBNR}
how the waveform mismatch vaguely follows a trend roughly
proportional to $e^{\Sperp}$.
Additionally, at least for the
precessing binaries used in this article, the values of $\Sperp$ 
are less degenerate than $\chi_p$, which we think would be a 
desirable property.

In summary:
\begin{enumerate}[label=(\roman*)]
 \item \TEOB{} matched \NRsurP{} to better than 0.99 for 74\%
 and better than 0.965 for 91\% of the \nNR{} cases with $\chi_p$ ranging up to $0.75$.
 Even for $\iota_s=\pi/2$, 85\% of the matches were greater than 0.965.
 \item For the same cases, 85\% of the \TEOB-\SEOBNR{} and
 77\% of the \TEOB-\texttt{IMRPhenomPv3HM} matches
 exceeded 0.965 with higher percentages for low-inclination matches,
 and lower ones for high inclinations.
 \item For the additional set consisting of 1030 binaries, 
 89\% of the $\iota_s=0,\pi/6$, \TEOB-\SEOBNR{} matches were greater than 0.965, which dropped to 73\% for $\iota_s=\pi/3$.
 \item Perhaps not surprisingly, the agreement between \TEOB{} and
 $\{$\NRsurP, \texttt{IMRPhenomPv3HM}, \SEOBNR$\}$ worsens for
 cases with stronger precession indicated by larger values of $\chi_p$ 
 (and $\Sperp$). Additionally, there is increasing disagreement for 
 binaries with large negative spins and small mass ratios.
 In particular, the three-dimensional region of the parameter space
 bounded roughly by $\chi_p \gtrsim 0.5, \chi_{\text{eff},0}\lesssim -0.3,
 q \lesssim 0.25$ has the densest population of matches less than 0.85.
 \end{enumerate}

The major cause of the disagreement is the mismatch of the non-precessing
modes. Any case for which the non-precessing (2,2) mode, 
$h_{22}^\text{NP}$, matches less than 0.965 will yield strain matches of 
$\lesssim 0.965$ as $h_{22}^\text{NP}$ contributes the most to the 
precessing (twisted) (2,2) mode 
which in turn is the dominant mode in the strain for most inclinations.
While there is only one non-precessing (2,2) mode match of less
than 0.99 between \TEOB{} and \NRsurP{} for the set of 200 binaries,
7\% of the 1230 \TEOB-\SEOBNR{} non-precessing (2,2) mode matches
are less than 0.965. These percentages increase to roughly 40\% and
42\% for the matches of $h_{21}^\text{NP}$ for the same sets above.
As $h_{21}^\text{NP}$'s contribution to the strain increases with respect to that of the $h_{22}^\text{NP}$'s
with increasing inclination, the mismatches of $h_{21}^\text{NP}$ 
affect the high-inclination cases more as confirmed by our findings.

One possible explanation for the increase in \TEOB-\NRsurP{} and
\TEOB-\SEOBNR{} mismatches with increasing $\chi_p$ is the fact
\TEOB{} twists \emph{constant}-spin, non-precessing waveforms, i.e., 
$\Sa(t)=\chi_1 m_1^2, \Sb(t)=\chi_2 m_2^2$, whereas both
\NRsurP{} and \SEOBNR{} twist so-called co-precessing waveforms with
time-varying $\Sa(t),\Sb(t)$ obtained either from fitting to NR data
or from the SEOB dynamics. 
Moreover, like \TEOB, \texttt{IMRPhenomPv3HM} also twists constant-spin 
waveforms and Ref.~\cite{Khan:2019kot} reports that the worst match against 
SXS NR simulations happens for a ``strongly precessing system'' with 
$\chi_p =0.78$ \cite{Pratten:2020ceb} and $q=1/6$.
Similarly, Ref.~\cite{Pratten:2020ceb} states that the worst \texttt{IMRPhenomXPHM} 
matches with respect to SXS simulations also occur for ``strongly precessing systems'' and $q\le 1/5$.
There is also Fig.~14 of Ref.~\cite{Ossokine:2020kjp}, where significant \SEOBNR-\texttt{IMRPhenomPv3HM} disagreement is observed for $q\lesssim 0.1, \chi_p\gtrsim 0.6$. Be that as it may, 
without a systematic study, our ``constant-spin-twist'' hypothesis
can not be tested, but we hope to do this after upgrading \TEOB{} as
we detail next.

Our most immediate task for the next version of \TEOB{}
is to add ringdown to the twisted modes.
One way to do this is as in Ref.~\cite{Babak:2016tgq}: 
by Euler-rotating the inspiralling modes
to the $\J_\text{peak}$ frame to attach the 
ringdown portion of the modes, where $\J_\text{peak}$ is extracted from the
solutions to the precession ODEs at a certain peak.
The stitched inspiral-merger-ringdown GW modes are then rotated
to the desired inertial frame.
It seems, however, that these steps might be redundant as \SEOBNR{}
successfully stitches the inspiral-merger-ringdown portions in the co-precessing frame \cite{Ossokine:2020kjp}.
See the end of Sec.~\ref{Sec:Twist} for a more detailed discussion.

The next task, after the incorporation of merger-ringdown,
is to add higher ($\ell \ge 3$) modes to \TEOB{}.
As Ref.~\cite{Nagar:2020pcj} states, the non-precessing 
\texttt{TEOBResumS} $(3,\pm3), (3,\pm2), (4,\pm4)$, and $(5,\pm5)$ modes show excellent agreement
with NR results, so they can be twisted then added to the strain. 
Thus, in principle, \TEOB{} can extend up to $\ell=5$, albeit in an incomplete manner,
but \SEOBNR{} also only has these modes [no $(3,\pm2)$] and has shown
improved agreement as compared to its ($\ell=2$)-only version \cite{Ossokine:2020kjp}.

Another planned improvement is to couple 
the precession equations to the \TEOBResumS{} dynamics.
This will enable us to generate aligned-spin waveforms with
time-varying $\chi_1=\Sa(t)\cdot \LN(t)/m_1^2, \chi_2=\Sb(t)\cdot \LN(t)/m_2^2$.
This upgrade might improve \TEOB's agreement with \NRsurP{} and
\SEOBNR{} for the strongly precessing cases.
Finally, we will test whether or not replacing the \STT{} 
expression for $\dot{v}$ with one obtained from
the aligned-spin \TEOBResumS{} dynamics may further improve \TEOB's performance.

As it stands, the current version of \TEOB{}
yields values greater than $0.965$ for $91\%,86\%\footnote{For the entire set
of 1230 cases and the inclinations considered here.}
, 77\%$ of the matches
with \NRsurP, \SEOBNR{}, and \texttt{IMRPhenomPv3HM} respectively. 
The significantly disagreeing
cases either have very strong precession, small mass ratios or
rather negative spins.
A nice feature of \TEOB{} is that
it is fast thanks to the post-adiabatic method implemented in \TEOBResumS{}
which ``rushes'' the inspiral \cite{Nagar:2018gnk}. \
We expect that, with the above additions, \TEOB{} will become
another useful precessing approximant for the analysis of future GW 
events. \TEOB{} will be added to the \TEOBResumS{} \verb|git| 
repository 
\url{https://bitbucket.org/eob_ihes/teobresums/wiki/Home}.

\begin{acknowledgments}
S.~A. acknowledges support from the University College Dublin Ad Astra Fellowship.
S.~A. and S.~B. acknowledge support by the EU H2020 under ERC Starting Grant, no.~BinGraSp-714626.
R.~G. acknowledges support from the Deutsche Forschungsgemeinschaft (DFG) under Grant No. 406116891 within the Research Training Group RTG 2522/1. 
S.~A. thanks Alessandro Nagar, Katerina Chatziioannou, Riccardo Sturani, Jonathan Thompson, and Marta Colleoni for helpful discussions.
S.~A. is also grateful to Niels Warburton for sending him files essential for this work
and to Eda Vurgun for her laptop during S.~A.'s self-exile in times of Covid-19.
This work makes use of the Black Hole Perturbation Toolkit 
\url{http://bhptoolkit.org/} and
the SimulationTools analysis package \url{http://simulationtools.org/}.
\end{acknowledgments}

\appendix
\section{Derivation of the post-Newtonian spin precession equations up to N4LO}
\label{sec:AppA}
This section builds upon the work of Ref.~\cite{Sturani_note}.
Recall that the $\Lhatdot$ equation is obtained by imposing total angular momentum conservation, $\dot{\mbf{J}}=0$ which leads to
\be
\dot{\mbf{\,L}} = - \Sadot - \Sbdot \label{eq:Ldot_eq}.
\ee
$\mbf{L}$ is provided up to 3.5\,PN in, e.g., Eq.~(4.7) of Ref.~\cite{Bohe:2012mr}
which we rewrite in the following compact form
\begin{align}
\mathbf{L} &= \f{\eta}{v}\left\{\Lhat \left[1 + v^2\left(\f{3}{2}+\f{\eta}{6}\right)\right. \right. \nn \\
&\hspace{19mm} \left. \,+ v^4\left( \f{27}{8}-\f{19\eta}{8}+\f{\eta^2}{24}\right)+ \ord(v^6)\right]\nn \\
&\hspace{11mm}  + v^3\Delta\mbf{L}_{1.5\text{PN}}^S + v^5\Delta\mbf{L}_{2.5\text{PN}}^S+v^7\Delta\mbf{L}_{3.5\text{PN}}^S\nn \\
&\hspace{8mm} \left. \textcolor{white}{\f{1}{2}}+\ord(v^8)\right\},\label{eq:Lvec}
\end{align}
where we defined the terms $\Delta\mbf{L}_{n\text{PN}}^S$ with $n=1.5, 2.5, 3.5$
with their explicit $v$ scalings factored out.
From Ref.~\cite{Bohe:2012mr}, we can extract
\begin{align}
\Delta\mbf{L}_{1.5\text{PN}}^S &= {\bm{\el}} \left(-\f{35}{6}S_\el -\f{5}{2}\delta m\, \Sigma_\el\right)\label{eq:L1p5PN} \\ 
& -{\bm{\lambda}} \left(3 S_\lambda +\delta m\, \Sigma_\lambda\right)
+{\mbf{n}} \left(\f{1}{2} S_n +\f{1}{2}\delta m\, \Sigma_n\right)\nn,
\end{align}
where $\elhatp = \Lhat$, $\nhatp=\mbf{r}/|\mbf{r}|$ is the relative separation unit vector, and
$\lhatp=\elhatp\times\nhatp$.
Moreover, $S_{\el,\la,n}\equiv \{\elhatp,\lhatp,\nhatp\}\cdot \mbf{S}, 
\Sigma_{\el,\la,n}\equiv \{\elhatp,\lhatp,\nhatp\}\cdot \bm{\Sigma} $,
where $\mbf{S}=\Sa+\Sb, \bm{\Sigma}=\Sb/m_2-\Sa/m_1$.
Defining $\mbf{S}_{1\el}\equiv \elhatp (\elhatp\cdot \Sa)$ and similarly for
$\mbf{S}_{1\la},\mbf{S}_{1n}$ as well as the $1\to2$ counterparts,
Eq.~\eqref{eq:L1p5PN} becomes
\begin{align}
\Delta\mbf{L}_{1.5\text{PN}}^S =&-\f{5}{6m_1}(3M+m_1)\mbf{S}_{1\ell}+\f{(M-m_1)}{2m_1}\mbf{S}_{1n}\nn\\ & -\f{(M+m_1)}{m_1}\mbf{S}_{1\la}
+(1\to 2) \label{eq:L1p5PN_v2},
\end{align}
where we restored $M=m_1+m_2$ for clarity in this section.
We can now orbit-average this expression using 
$\langle \hat{n}^i \hat{n}^j\rangle = \langle \hat{\la}^i \hat{\la}^j\rangle
=\tfrac{1}{2}(\delta^{ij}-\hat{\el}^i\hat{\el}^j) $ which yields
$\langle \mbf{S}_{1 n}\rangle = \langle\mbf{S}_{1\la}\rangle =\tfrac{1}{2}(\Sa-\mbf{S}_{1\el})$.
Substituting these orbit-average terms into Eq.~\eqref{eq:L1p5PN_v2} we arrive at
\begin{align}
 \Delta\mbf{L}_{1.5\text{PN}}^S =&-\f{M+3m_1}{4m_1}\Sa-\f{(27M+m_1)}{12m_1} \Lhat (\Lhat\cdot \Sa)
 \nn\\ &+(1\to 2) \label{eq:L1p5PN_v3}.
\end{align}

Similarly, with some more determination, one can obtain
\begin{align}
 &\Delta\mbf{L}_{2.5\text{PN}}^S =\left(\f{7M-31m_1}{16m_1}+\eta \f{22M+9m_1}{48m_1} \right)\Sa\nn\\ 
 &+\left[-\f{49M+39m_1}{16m_1}+\eta\left(\f{59M}{24m_1}-\f{13}{144} \right) \right]
 \Lhat (\Lhat\cdot \Sa)\nn\\
 &+ (1\to 2) \label{eq:L2p5PN}.
\end{align}
Eq.~(\ref{eq:L1p5PN}) inside Eq.~\eqref{eq:Lvec}
together with Eqs.~(\ref{eq:S1dot_N4LO},\,\ref{eq:S2dot_N4LO}) give us all the pieces
that we need to go to N4LO [Eq.~\eqref{eq:L2p5PN} enters at N5LO so we drop it.]
For clarity, let us once again consider NNLO first.
At this order, Eq.~\eqref{eq:Lvec} becomes
\begin{align}
\mbf{L} =& \Lhat\, \f{\eta}{v}\text{L}_{1\text{PN}} + \eta v^2 \left(c_{S1} \Sa+c_{S2}\Sb\right) \nn\\
+ & \eta v^2\Lhat \left(c_{S1L} \Lhat\cdot\Sa+c_{S2L}\Lhat\cdot\Sb\right)\label{eq:Lvec_NNLO},
\end{align}
where $\text{L}_{1\text{PN}}\equiv 1+v^2\left(\tfrac{3}{2}+\tfrac{1}{6}\eta\right) $ and 
the constants $c_{S1}, c_{S1L}$, etc., are given in Eqs.~(\ref{eq:c_S1},\,\ref{eq:c_S1L}).
Differentiating Eq.~\eqref{eq:Lvec_NNLO} with respect to time, we obtain
\begin{align}
\Lhatdot^\text{NNLO} = \f{v}{\eta}\f{1}{\text{L}_{1\text{PN}}}& \left[ -\Sadot^\text{NNLO}-\Sbdot^\text{NNLO} \right. \nn\\ 
& \left. -\eta v^2\left(c_{S1} \Sadot^\text{LO}+c_{S2}\Sbdot^\text{LO}\right)\right] \label{eq:LNhatdot_NNLO},
\end{align}
where, e.g., $\Sadot^\text{NNLO}$ implies that only terms that scale as $v^{\le 7}$ should be retained.
Several simplifications occurred in reaching Eq.~\eqref{eq:LNhatdot_NNLO}.
First, the second $\Sadot, \Sbdot$ terms contribute only at the LO. 
This is because of the factor of $v^2$ in front,
which means that at our required order, i.e., NNLO, 
the terms multiplying $v^2$ can be at most $\propto v^5$ 
which is LO for $\Sadot,\Sbdot$ as can be seen from Eqs.~(\ref{eq:S1dot_NLO}, \ref{eq:S2dot_NLO}).
Second, all the $ c_{S1L}, c_{S2L}$ terms have dropped from Eq.~\eqref{eq:LNhatdot_NNLO}
because (i) $v^2 \Lhatdot \propto v^8$, i.e., is N3LO
and (ii) at NNLO only $v^2 \Lhat\cdot \Sadot^\text{LO}$ scales as $v^7$, but is actually zero because $\Lhat\perp \Sadot^\text{LO}$ as is clear from Eqs.~(\ref{eq:S1dot_NLO}, \ref{eq:S2dot_NLO}).
\begin{widetext}

\begin{figure}[t]
\hspace{-0.9cm}
   \includegraphics[width=1.0\textwidth]{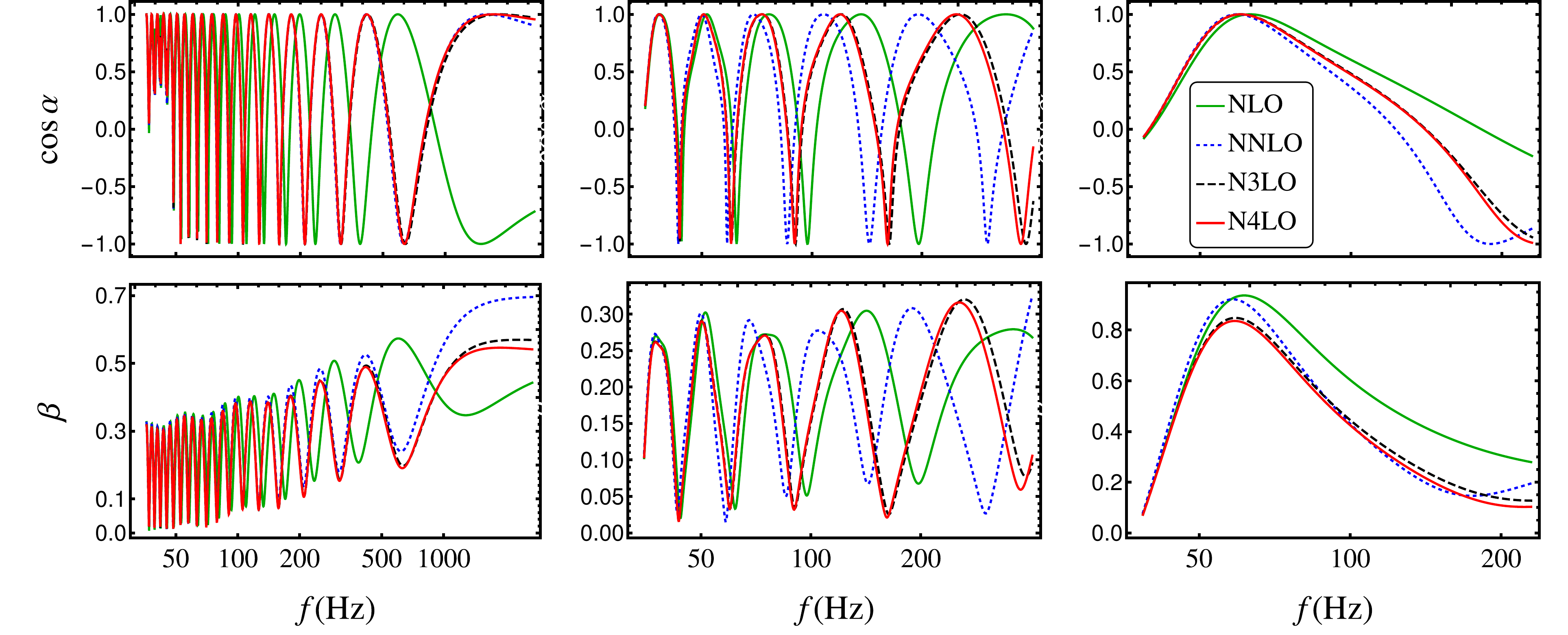}
\caption{\label{fig:Euler_angles} 
The spherical angles of $\LN(t)$ in the $\Livec$ frame described in Sec.~\ref{sec:Frames}
for three separate cases: binary neutron star (left panels), black hole neutron star (middle panels), and binary black hole (right panels) inspirals.
$\alpha$ is the azimuthal angle and $\beta$ is the polar angle (see Fig.~\ref{fig:Frames}).
In each panel, we show the angle obtained from solving the precession ODEs truncated
at four different orders: NLO, NNLO, N3LO, and N4LO.
From left to right, the binaries respectively have $M=3,20, 35 \Msun$, 
$q\approx 0.85, 0.11,0.35$, and $\chi_p\approx 0.78,0.22,0.74$.
$f$ represents the (2,2)-mode gravitational wave frequency.
}
\end{figure}
\end{widetext}

Pushing now to N4LO, Eq.~\eqref{eq:Ldot_eq} becomes
\begin{align}
 \dot{\mbf{\,L}}&=\Lhatdot \f{\eta}{v}\text{L}_{2\text{PN}} +\eta v^2 \dot{\Delta\L}^S_{1.5\text{PN}}
 = -\Sadot^\text{N4LO}-\Sbdot^\text{N4LO}
 \label{eq:Ldot_eq2},
\end{align}
where $\text{L}_{2\text{PN}} $ is given in Eq.~\eqref{eq:L_2PN}.
Note that we omit the radiation reaction
terms starting at NNLO via $\eta \dot{v}/v^2 \propto v^7$ in $\dot{\mbf{\,L}}$ 
because they drop out from $\LhatdotNew$ given in Eq.~\eqref{eq:Lperpdot_N4LO}
since these terms are all parallel to $\Lhat$.
The effects of radiation reaction are incorporated via $v=v(t)$ in the precession ODEs
after the standard change of variables $d/dt \to \dot{v}(v)d/dv $ in Eqs.~(\ref{eq:S1dot_NLO}) -
(\ref{eq:S1dot_NNLO}).

Explicitly writing out Eq.~\eqref{eq:Ldot_eq2} at N4LO then rearranging gives us 
Eq.~\eqref{eq:LNhatdot_N4LO},
where we used the property that $\dot{\mbf{S}}_i \perp \mbf{S}_i$  up to NLO.
%
%
%
In terms of powers of $v$, each term in Eq.~\eqref{eq:LNhatdot_N4LO}
goes up to $ v^9$, i.e., N4LO as defined.

We can now obtain $\LN$, therefore, the angles $\alpha$ and $\beta$ at any order of our choosing varying from NLO to N4LO,
which we show in Fig.~\ref{fig:Euler_angles} as functions of the (2,2)-mode GW frequency
for three different precessing compact binary inspirals. 
As can be seen in the figure, the angles from different orders remain very close to each other in general
until the binaries enter their respective strong-gravity regimes. 
The angle dephasing between different orders happens earlier and is most prominent for the most asymmetric system in the figure,
i.e., a black hole neutron star binary with $M=20\Msun$ and $q\approx 0.11$.
The differences between the N3LO and N4LO angles are much smaller,
expectedly so since the differences of these two orders scales as $v^9$.

A thorough survey of the effects of the truncation order of the precession ODEs,
the instantaneous terms (entering at N3LO), 
and the neglected terms would be beneficial to the entire gravitational-wave community.
Ref.~\cite{Ossokine:2015vda} has already done some work in this regard,
but a systematic, large-scale analysis quantified in terms of consequences to parameter estimation
remains to be undertaken at this point.


\section{Results of using NLO angles and a different twist formula}\label{sec:AppB}

In this section, we briefly show results from two additional test we conducted:
\begin{inparaenum}[(1)]
 \item Using Euler angles in the twist formula \eqref{eq:hlm_Twist2} that are obtained
 from the precession ODEs truncated at NLO as given in Eqs.~\eqref{eq:S1dot_NLO}-\eqref{eq:LNdot_NLO}.
 \item Using N4LO Euler angles in an alternate twist formula.
 Specifically, we have chosen to test the expression provided by
 Eq.~(A2) of Ref.~\cite{Khan:2019kot}
 \end{inparaenum}
\be
h_{\ell m}^\text{T}(t) = e^{i m \alpha}\sum_{m'=-l}^l e^{-i m'\gamma} d^\ell_{m',m}(-\beta)\, h_{l m'}^\text{NP}\label{eq:hlm_Twist_Khan} \, . 
\ee
This version differs from our twist formula \eqref{eq:hlm_Twist2} in the signs of the $\alpha$ and $\gamma$ exponents. 
For convenience, we redisplay our expression 
\be
h_{\ell m}^\text{T}(t) = e^{-i m \alpha}\sum_{m'=-l}^l e^{i m'\gamma} d^\ell_{m',m}(-\beta)\, h_{l m'}^\text{NP}\label{eq:hlm_Twist3} \, .
\ee
\begin{figure}[t]
\centering   
   \includegraphics[width=0.49\textwidth]{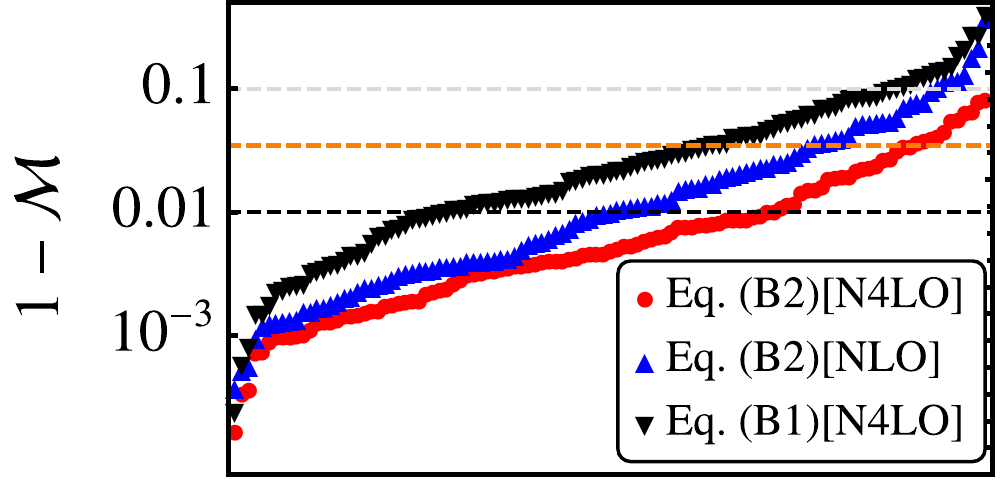}
\caption{\label{fig:Alt_twists} 
Performance of two alternate \TEOBResumS{} twists against our standard twist measured in terms of 
$\el=2$ detector strain mismatches with \NRsurP{} (50 cases) and \SEOBNR{} (60 cases)
ordered by increasing mismatch.
Red circles represent mismatches obtained using our standard expression,
dubbed Eq.~\eqref{eq:hlm_Twist3}[N4LO]:
twisting via Eq.~\eqref{eq:hlm_Twist3} with angles obtained from the precession ODEs truncated
at N4LO. Similarly, the blue triangles represent mismatches obtained with the
same twist formula, but with angles coming from the NLO-truncated ODEs, hence dubbed
Eq.~\eqref{eq:hlm_Twist3}[NLO]. Finally, the inverted black triangles represent mismatches
resulting from using the alternate twist formula \eqref{eq:hlm_Twist_Khan}
at N4LO.
As before, the horizontal dashed black, orange, and gray lines mark 
$\M=0.99, 0.965, 0.9$.
It is clear from the figure that our standard expression produces the best matches.
We left the horizontal axis unlabelled since we reordered the total of 110 cases 
in terms of increasing $1-\M$.}
\end{figure}

For both tests, we used a subset of precessing compact binary inspirals 
that is a combination of 50 cases from our \NRsurP{} set 
and 60 cases from our \SEOBNR{} set.
Using Eq.~\eqref{eq:hlm_Twist_Khan} at N4LO and Eq.~\eqref{eq:hlm_Twist3} at NLO
we generated two new sets of twisted $\el=2$ \TEOBResumS{} modes with which
we then computed the $\el=2$ detector strain matches as before.
We show how these two alternate twists perform against ours,
dubbed Eq.~\eqref{eq:hlm_Twist3}[N4LO], in Fig.~\ref{fig:Alt_twists},
where it is evident that our twist produces consistently the smallest mismatches (red circles).
The alternate twist formula of Eq.~\eqref{eq:hlm_Twist_Khan} 
is clearly the worst choice
producing $\M>0.965$ for only about two thirds of the set (black inverted triangles).
The reason why Eq.~\eqref{eq:hlm_Twist_Khan}[N4LO] still somehow manages
to mostly yield $\M>0.965$ is due to both the fact that 
$\gamma$ remains close to $\alpha$ because $\beta$, starting from zero, is small for most binaries,
and that the twisted $(2,\pm2)$ modes differ by a small amount. 
Therefore, in binaries for which 
$\beta(t)\ll 1$ and the precessing $(2,\pm2)$
modes dominate the mode-sum in the strain formula \eqref{eq:strain}, Eqs.~\eqref{eq:hlm_Twist_Khan}
and \eqref{eq:hlm_Twist3} are nearly equal under the $m\to -m$ exchange, 
thus produce twisted waveform strains that are very close to each other.

Returning to Fig.~\ref{fig:Alt_twists}, we see that
the NLO version of our twist 
performs somewhat well in the sense that roughly three quarters
of the cases yielded $\M >0.965$ (blue triangles).
The details of the differences in the plotted NLO, N4LO mismatches
lay with the differences in the Euler angles used in the respective twists.
We have already shown in Fig.~\ref{fig:Euler_angles} how these Euler angles vary
as the ODE truncation order goes from NLO to N4LO. 
For most cases, the difference in the angles become significant only in the last few orbital
cycles, corresponding to the small differences between the NLO and N4LO mismatches of Fig.~\ref{fig:Alt_twists}.
But for cases with small $q$, the differences in the Euler angles becomes more significant
as can be seen in the middle panels of Fig.~\ref{fig:Euler_angles}.
It is possible that the speed-up gained in using NLO-truncated precession ODEs,
instead of N4LO, is significant enough to justify their use in parameter estimation.
As we have not yet carried out detailed speed tests of our code, we can not verify
or refute this hypothesis,
but will do so with the next version of \TEOB.

\bibliography{references,local_bib} 

\end{document}